\documentclass[11pt,prd,aps,nofootinbib,preprintnumbers,tightenlines]{revtex4}

\usepackage{epsf,graphicx}
\usepackage{times,latexsym,euscript,amstext,amssymb,amsbsy,amsmath,amsfonts}
\usepackage{epsfig}

\usepackage[T1]{fontenc}
\usepackage{charter}
\usepackage{wrapfig}
\usepackage[a4paper,portrait,top=1.5cm, bottom=2.2cm,left=1.5cm,right=1.5cm]{geometry}
\usepackage[colorlinks,citecolor=blue,linktoc=all,linkcolor=cyan]{hyperref}
\def\rmqp{{ \hbox{\rm q}'} }

\def\calm{{\mathcal M}}
\def\qt0{\tilde{q}_0}

\def\b0{\boldsymbol{0}}

\def\vare{\varepsilon}

\newcommand{\be}{\begin{equation}}
\newcommand{\ee}{\end{equation}}
\newcommand{\bea}{\begin{eqnarray}}
\newcommand{\eea}{\end{eqnarray}}
\newcommand{\nn}{\nonumber}
\def\eqlab#1{\label{eq:#1}}

\def\barr{\left(\begin{array}{c}}
\def\earr{\end{array}\right)}
\def\bmat{\left(\begin{array}{cc}}
\def\emat{\end{array}\right)}

\DeclareMathOperator{\re}{Re}

\begin{document}
\preprint{MITP/16-119}

\title{Spin-dependent sum rules connecting real and virtual Compton scattering verified}

\author{Vadim Lensky}
\affiliation{Institut f\"ur Kernphysik, Cluster of Excellence PRISMA, Johannes Gutenberg Universit\"at, Mainz D-55099, Germany}
\affiliation{Institute for Theoretical and Experimental Physics, 117218 Moscow, Russia}
\affiliation{National Research Nuclear University MEPhI (Moscow Engineering Physics Institute), 115409 Moscow, Russia}
 \author{Vladimir Pascalutsa}
\author{Marc Vanderhaeghen}
\affiliation{Institut f\"ur Kernphysik, Cluster of Excellence PRISMA, Johannes Gutenberg Universit\"at, Mainz D-55099, Germany}

\author{Chung Wen Kao}
\affiliation{Department of Physics and Center for High Energy Physics, Chung-Yuan Christian University, Chung-Li 32023, Taiwan}

\date{\today}

\begin{abstract}
We present a detailed derivation of the two sum rules
relating the spin polarizabilities measured
in real, virtual, and doubly-virtual Compton scattering. For example, 
the  polarizability $\delta_{LT}$, accessed in inclusive electron
scattering, is related to the spin polarizability $\gamma_{E1E1}$
and the slope of generalized polarizabilities $P^{(M1,M1)1}-P^{(L1,L1)1}$,
measured in, respectively, the real and the virtual Compton scattering.
 We verify these sum rules
in different variants of chiral perturbation theory,
discuss their empirical verification for the proton, and prospect 
their use in studies of the nucleon spin structure.
\end{abstract}

\maketitle
\tableofcontents
\newpage
\section{Introduction}
The low-energy nucleon structure  is presently at the forefront of  
many precision studies of the Standard  Model and beyond.  Given
the complexity  of low-energy  QCD, a popular method 
of calculating the nucleon-structure  effects is  the data-driven
approach based on model-independent relations, such as sum rules.
Perhaps the best known sum rule  for the electromagnetic
structure of the nucleon is  the Gerasimov-Drell-Hearn (GDH)
sum 
rule~\cite{GellMann:1954db, Gerasimov:1965et, Drell:1966jv}, relating the  anomalous magnetic moment of the nucleon to a weighted integral over the polarized photo-absorption cross section. 
An even older sum rule is the one of Baldin~\cite{Baldin}, 
which gives the sum of the electric $\alpha_{E}$ and magnetic $\beta_{M}$
dipole polarizabilities in terms
the total cross section $\sigma(\nu)$ as follows:
\begin{equation}
\alpha_{E}+\beta_{M}=\frac{1}{2\pi^2}\int\limits^\infty_{\nu_0}
\frac{\mathrm{d}\nu}{\nu^2}\,\sigma(\nu)\,,
\end{equation}
where $\nu$ is the photon energy in the laboratory frame, 
and $\nu_0$ is the inelastic threshold.  This  sum rule  has
proven to be very useful for an accurate extraction of  
the nucleon  polarizabilities, see Refs.~\cite{Drechsel:2002ar, Schumacher:2005an, Griesshammer:2012we, Hagelstein:2015egb} for reviews.

The Baldin sum rule, derived from general considerations of the
forward Compton scattering amplitude, is easily generalized to the case of virtual photons, i.e., the forward double-virtual Compton scattering (VVCS). The sum of the two
polarizabilities becomes dependent on the photon virtuality $Q^2$,
and is connected with the unpolarized nucleon structure function $F_1$, measured in electron-nucleon
scattering, via
\begin{equation}
\alpha_{E}(Q^2)+\beta_{M}(Q^2)=\frac{8\alpha_\mathrm{em} M}{Q^4}\int\limits_0^{x_0}
\mathrm{d}x\, x\, F_1(x,Q^2)\,,
\end{equation}
where $M$ is the nucleon mass, $\alpha_\mathrm{em}$ is the fine structure constant,
and $x=Q^2/2M\nu$ is the Bjorken scaling variable with $x_0$ corresponding to the inelastic
threshold. Other forward sum rules involving the spin dependent nucleon structure functions allow for such a generalization~\cite{Anselmino:1988hn, Ji:1999mr}, too. In this way, one can characterize the spin-dependent VVCS process through $Q^2$-dependent transverse ($\gamma_0$) or 
transverse-longitudinal ($\delta_{LT}$) spin polarizabilities of the nucleon~\cite{Drechsel:2000ct,Drechsel:2002ar,Drechsel:2004ki}. 
These polarizabilities are thus related with the nucleon spin-dependent structure functions, and have been the subject of dedicated experimental activities at the Jefferson Lab, 
see~\cite{Kuhn:2008sy, Chen:2010qc} for reviews. 

The nucleon's response to external electromagnetic fields can also be probed through the 
virtual Compton scattering (VCS) process, in which 
the initial photon has finite virtuality $Q^2$ whereas the final one
is real. 
The linear response of a nucleon in the low-energy VCS process can be expressed through generalized polarizabilities (GPs)~\cite{Guichon:1995pu}, see Ref.~\cite{Guichon:1998xv}
for a detailed review. 
The GPs, which encode the spatial distribution of the polarization densities in a nucleon~\cite{Gorchtein:2009qq}, 
have been the subject of several dedicated experiments  at MAMI~\cite{Roche:2000ng,Janssens:2008qe,d'Hose:2006xz,Doria:2015dyx,Correa:thesis}, MIT-Bates~\cite{Bourgeois:2006js,Bourgeois:2011zz}, 
and JLab~\cite{Laveissiere:2004nf,Fonvieille:2012cd}. 

As the VCS process is a non-forward process and apparently asymmetric under photon crossing, it has precluded 
an immediate connection, via sum rules, of GPs with  photoabsorption cross sections. Nonetheless,  a new type of relation, presented
recently in  Ref.~\cite{Pascalutsa:2014zna}, allows to 
relate two of the spin-dependent GPs with quantities measured in RCS and VVCS. The new relations provide an extension  of the GDH sum rule to finite virtuality, and as a result involve new
quantities which are  accessible in independent experiments. 

In this work we provide a detailed derivation of the two new sum rules. We first discuss the forward double virtual Compton scattering (Section~\ref{sec2}), its low-energy expansions (Section~\ref{sec3}), and the derivation of forward sum rules for the four amplitudes 
characterizing the VVCS on the nucleon (Section~\ref{sec4}). 
We then show how these sum rules are satisfied 
in heavy baryon chiral perturbation theory (Section~\ref{sec5}) 
as well as in its covariant counterpart --- baryon chiral perturbation theory (Section~\ref{sec6}).   
We discuss the phenomenological status 
of these sum rules and further experimental opportunities 
in Section~\ref{sec7}. In particular,
by using the sum rules, we will obtain an empirical prediction for the slope of one of the VCS response functions, denoted by $P_{TT}$, and will compare it with the dispersive
evaluations and with the predictions of baryon chiral perturbation theory calculations (Section~\ref{sec8}). Finally, we will present our conclusions in Section~\ref{sec9}.

\section{Sum rule derivation}

\subsection{Forward double virtual Compton scattering}
\label{sec2}

Our starting point is the double-virtual Compton on a nucleon
\bea
\gamma^\ast (q, \lambda) + N(p, s) \to \gamma^\ast(q^\prime, \lambda^\prime) + N(p^\prime, s^\prime),
\label{doublevcs}
\eea
where $\lambda, \lambda^\prime$ denote the photon helicities, 
and  $s, s^\prime$ are the nucleon helicities. 
This process is described by 18 helicity amplitudes introduced as:
\bea
T_{\lambda^\prime s^\prime, \lambda s} \equiv e^2
\varepsilon_\mu(q, \lambda) \varepsilon_\nu^\ast(q^\prime, \lambda^\prime) 
\, \bar u(p^\prime, s^\prime) M^{\mu \nu} u(p, s), 
\eea 
where $e$ is the proton electric charge, and 
$\varepsilon_\mu$ ($\varepsilon_\nu^\ast$) stands for the initial (final) nuclear polarization vector. 
The double virtual Compton tensor $M^{\mu \nu} $ can be Lorentz-decomposed as  
(in the notation of Ref.~\cite{DKK98}):
\bea
M^{\mu \nu} &=& 
\sum_{i \in J} B_i(q^2, q^{\prime \, 2}, q \cdot q^\prime, q \cdot P) \, T_i^{\mu \nu} , 
\nonumber  \\
&&J = \{ 1,..., 21\} \backslash \{5, 15, 16 \},
\label{eq:dvcs}
\eea
where $P = \frac{1}{2} ( p + p^\prime)$. 
The 18 independent tensors $T_i^{\mu \nu}$ can be constructed to be gauge invariant, 
and free of kinematical singularities as shown by Tarrach~\cite{Tarrach:1975tu}. 
The invariant amplitudes $B_i$ have definite transformation properties with respect to the photon crossing, as well as 
charge conjugation combined with nucleon crossing~\cite{DKK98}. The latter reference
also contains the low-energy expansions of $B_i$'s up to ${\cal O}(k^3) \, (k = \{q, q^\prime \})$, which 
will be useful in the following.   

To derive the sum rules one considers the {\it forward} double VCS process (VVCS), 
which is a special case of the process (\ref{doublevcs}), with $q^\prime = q$ and $p^\prime = p$. 
The VVCS process is described by the four invariant amplitudes, denoted by $T_1, T_2, S_1, S_2$, which are functions 
of $Q^2 \equiv -q^2$ and $\nu \equiv p \cdot q / M$. 
Its covariant tensor structure can be written as:
\bea
\label{vvcs}
\alpha_\mathrm{em} \, M^{\mu \nu}(\nu,Q^2)&=&
- \left \{ \left( -g^{\mu\nu}+\frac{q^{\mu}q^{\nu}}{q^2}\right)
T_1(\nu, Q^2) +\frac{1}{M^2} \left(p^{\mu}-\frac{p\cdot
q}{q^2}\,q^{\mu}\right) \left(p^{\nu}-\frac{p\cdot
q}{q^2}\, q^{\nu} \right) T_2 (\nu, Q^2) \right .\nn \\
&& \left . +\frac{i}{M}\,\epsilon^{\nu\mu\alpha\beta}\,q_{\alpha}
s_{\beta}\, S_1(\nu, Q^2) + \frac{i}{M^3}\,\epsilon^{\nu\mu\alpha\beta}\,q_{\alpha}
(p\cdot q\ s_{\beta}-s\cdot q\ p_{\beta})\, S_2 (\nu, Q^2)  \right\},
\eea
where the fine-structure constant $\alpha_\mathrm{em} \equiv e^2 / 4 \pi \simeq 1/137$ is conventionally introduced in defining the forward amplitudes $T_1$, $T_2$, $S_1$, and $S_2$. 
Furthermore, $\epsilon_{0123} = +1$, and $s^\alpha$ is the nucleon covariant
spin vector satisfying $s \cdot p$ = 0, $s^2 = -1$.  
The optical theorem relates the imaginary parts of the four
amplitudes appearing in Eq.~(\ref{vvcs}) to the four structure functions of inclusive electron-nucleon scattering as:
\bea
\label{optical}
{\rm{Im}}\ T_1(\nu,\,Q^2) = \frac{e^2}{4 M} F_1(x,\,Q^2) \,&,& \quad
{\rm{Im}}\ T_2(\nu,\,Q^2)  = \frac{e^2}{4 \nu}  F_2(x,\,Q^2) \, , \nn \\
{\rm{Im}}\ S_1(\nu,\,Q^2)  = \frac{e^2}{4 \nu}  g_1(x,\,Q^2) \, &,& \quad
{\rm{Im}}\ S_2(\nu,\,Q^2)  = \frac{e^2}{4} \frac{M}{\nu^2} g_2(x,\,Q^2) \, ,
\eea
where $x \equiv Q^2 / 2 M \nu$, and where 
$F_1, F_2, g_1, g_2$ are the conventionally defined structure functions which parametrize 
inclusive electron-nucleon scattering. 
The imaginary parts of the forward scattering amplitudes, Eqs.~(\ref{optical}),
get contributions from both elastic scattering at $\nu = \nu_B  \equiv Q^2/(2M)$ or equivalently $x=1$, as well as 
from inelastic processes above the pion threshold, corresponding with $\nu > \nu_0 \equiv m_\pi + (Q^2 + m_\pi^2)/(2 M)$ 
or equivalently $x < x_0 \equiv Q^2 / (2 M \nu_0)$. 
The elastic contributions are obtained as pole parts of the direct and crossed nucleon Born diagrams. 
The latter are conventionally separated off the Compton scattering tensor in order to define 
structure-dependent constants, such as polarizabilities. The Born terms are defined by 
using the electromagnetic vertex for the transition $\gamma^* (q) + N(p) \to N(p + q)$ as given by
\be
\label{emvertex}
\Gamma^\mu \;=\; F_D(Q^2) \, \gamma^\mu \;+\;
F_P(Q^2) \, i \sigma^{\mu \nu} \frac{q_\nu}{2 M} \, ,
\ee
with $F_D$ and $F_P$ the Dirac and Pauli form factors of nucleon $N$,
normalized to $F_D(0)=e_N$ and $F_P(0)=\kappa_N$,
where $e_N$ is the charge in units of $e$, and where $\kappa_N$ is 
the anomalous magnetic moment in units of $e/2M$; $\sigma^{\mu\nu} = (i/2)[\gamma^\mu,\gamma^\nu]$.
This choice of the electromagnetic vertex
ensures that the Born contributions are gauge invariant and leads to
the following contributions:
\bea
\label{born}
T_1^\mathrm{Born}  & = &
 -  \frac{\alpha_\mathrm{em}}{M}
(F_D^2 + \frac{\nu_B^2}{\nu^2-\nu_B^2+i \vare} \,G_M^2) \,, \hspace{0.9cm}
T_2^\mathrm{Born}  =
- \frac{\alpha_\mathrm{em}}{M} \,\frac{Q^2}{\nu^2-\nu_B^2+i \vare} \,
(F_D^2 + \tau \, F_P^2 ) \, ,\nn \\
S_1^\mathrm{Born} & = &
 -  \frac{\alpha_\mathrm{em}}{2M}
(F_P^2 + \frac{Q^2}{\nu^2-\nu_B^2+i \vare}\,F_D G_M ) \,, \hspace{0.5cm}
S_2^\mathrm{Born}  =
\frac{\alpha_\mathrm{em}}{2}\frac{\nu}{\nu^2-\nu_B^2+i \vare}
\,F_P G_M \, ,
\eea
with $G_M\,(Q^2) = F_D\,(Q^2)+F_P(Q^2)$, 
and $\tau \equiv Q^2/ 4M^2$. 
The Born contributions of Eq.~(\ref{born}) can be split into non-pole 
and pole contributions in a dispersion relation framework.  The  
pole contributions (also called elastic contributions) can be immediately read off Eqs.~(\ref{born}). Their real parts 
are given by
\bea
\label{pole}
\re T_1^\mathrm{pole}  & = &
 -  \frac{\alpha_\mathrm{em}}{M} \frac{\nu_B^2}{\nu^2-\nu_B^2} \,G_M^2 \,, \hspace{2cm}
\re  T_2^\mathrm{pole}  =
- \frac{\alpha_\mathrm{em}}{M} \,\frac{Q^2}{\nu^2-\nu_B^2} \,
(F_D^2 + \tau \, F_P^2 ) \, ,\nn \\
\re  S_1^\mathrm{pole} & = &
 -  \frac{\alpha_\mathrm{em}}{2M} \frac{Q^2}{\nu^2-\nu_B^2}\,F_D G_M  \,, \hspace{0.75cm}
\re \left( \nu S_2^\mathrm{pole} \right)  =
\frac{\alpha_\mathrm{em}}{2}\frac{\nu_B^2}{\nu^2-\nu_B^2} \,F_P G_M \, ,
\eea

\subsection{Low-energy expansions}
\label{sec3}

Following Ref.~\cite{DKK98}, in order to obtain a low-energy expansion (LEX) in $k = \{\nu, Q \}$ for the forward VCS amplitudes $T_1, T_2, S_1$, and $S_2$, we express them in terms of the $B_i$ of  Eq.~(\ref{eq:dvcs}):
\begin{subequations}
\bea
T_1(\nu, Q^2) &=& \alpha_\mathrm{em} \, \left\{ Q^2 \, B_1 - 4 M^2 \nu^2 \, B_2 + Q^4 \, B_3 - 4 M \nu Q^2 \, B_4 \right\}, \label{rel1} \\
T_2(\nu, Q^2) &=& \alpha_\mathrm{em} \, 4 M^2 Q^2 \left\{ - B_2 - Q^2 B_{19} \right\}, \label{rel2} \\
S_1(\nu, Q^2) &=& \alpha_\mathrm{em} \, M \left\{ - 4 M \nu \, B_7 + Q^2 \left[ B_8 + M (4 B_{10} + 2  B_{21}) + 4 B_{18} \right] \right\}, \label{rel3} \\
S_2(\nu, Q^2) &=& \alpha_\mathrm{em} \, M^2 \left\{- \frac{Q^2}{2}  B_6 - 2 B_{17} + M \nu  ( 4 B_{10} + 2  B_{21} ) - Q^2 \,B_{12} \right\}, 
\label{rel4}
\eea
\end{subequations}
where the $B_i$ also depend on $\nu$ and $Q^2$ for forward kinematics. 
We can next use the expansions  in $k = \{\nu, Q \}$ established in~\cite{DKK98}:
\bea
B_i &=& b_{i,0} + {\cal  O}(k^2), \quad \quad (i = 1, 2, 3, 8, 10, 18, 19, 21), \\
B_i &=& 2 M \nu \left[ b_{i,1} + {\cal  O}(k^2) \right] , \quad \quad (i = 4, 6, 7, 12, 17), 
\eea
where $b_{i,0}$ and $b_{i,1}$ are low-energy constants.
As we are only interested in the lowest-order terms in $k = \{\nu, Q \}$, we obtain 
the following LEXs for Eqs.~(\ref{rel1})--(\ref{rel4}):
\begin{subequations}
\bea
T_1(\nu, Q^2) &=& \alpha_\mathrm{em} \, \left\{ Q^2 \, b_{1, 0} - 4 M^2 \nu^2 \, b_{2, 0} + {\cal O}(k^4) \right\}, \label{lex1} \\
T_2(\nu, Q^2) &=& - \alpha_\mathrm{em} \, \left\{ 4 M^2 Q^2 \, b_{2, 0}  + {\cal O}(k^4) \right\}, \label{lex2} \\
S_1(\nu, Q^2) &=& \alpha_\mathrm{em} \, M \left\{ - 8 M^2 \nu^2 \, b_{7,1} + Q^2 \left[ b_{8,0} + M (4 \, b_{10,0} + 2 \, b_{21,0}) + 4 \, b_{18,0} \right] + {\cal O}(k^4)  \right\}, \label{lex3} \\
S_2(\nu, Q^2) &=& \alpha_\mathrm{em} \, M^3 \nu \left\{ - 4 \, b_{17,1} +  4 \, b_{10,0} + 2 \,  b_{21,0} 
+ {\cal O}(k^2)  \right\}.
\label{lex4}
\eea
\end{subequations}
Of the eight coefficients appearing in Eqs.~(\ref{lex1})--(\ref{lex4}), six can be related to the 
scalar and spin dipole polarizabilities as measured in real Compton scattering (RCS). 
As polarizabilities are conventionally defined by separating off the Born parts of the amplitudes, 
one splits the amplitudes into Born and non-Born parts as $T_1 = T_1^\mathrm{Born} + T_1^\mathrm{nB}$, and analogously for the other three amplitudes.  The Born parts are given by Eqs.~(\ref{born}).
The non-Born ($\mathrm{nB}$) parts of six of the low-energy constants are then 
expressed in terms of polarizabilities:
\begin{subequations}
\bea
b_{1, 0}^\mathrm{nB} &=& \frac{1}{\alpha_\mathrm{em}} \beta_M , 
\label{b1c0} \\
b_{2, 0}^\mathrm{nB} &=& - \frac{1}{\alpha_\mathrm{em}} \frac{1}{4 M^2} (\alpha_E + \beta_M) , 
\label{b2c0}  \\
b_{7, 1}^\mathrm{nB} &=& - \frac{1}{\alpha_\mathrm{em}} \frac{1}{8 M^2} \gamma_0 , 
\label{b7c1}  \\
b_{10, 0}^\mathrm{nB} &=& \frac{1}{\alpha_\mathrm{em}} \frac{1}{4 M} (\gamma_{M1 E2} + \gamma_{E1 M2} ) , 
\label{b10c0}  \\
b_{17, 1}^\mathrm{nB} &=& \frac{1}{\alpha_\mathrm{em}} \frac{1}{4 M} (\gamma_{M1 E2} - \gamma_{M1 M1} ) , 
\label{b17c1}  \\
b_{18, 0}^\mathrm{nB} &=& -\frac{1}{\alpha_\mathrm{em}} \frac{1}{2} \gamma_{M1 E2}, 
\label{b18c0} 
\eea
\end{subequations}
where $\alpha_E$ ($\beta_M$) are the electric (magnetic) dipole polarizabilities respectively, and 
$\gamma_{M1 E2}$, $\gamma_{E1 M2}$, $\gamma_{M1 M1}$, $\gamma_{E1 E1}$ are the lowest-order spin polarizabilities of the nucleon, which are related to the forward spin polarizability $\gamma_0$ as:
\bea
\gamma_0 = - \left( \gamma_{M1 E2} + \gamma_{E1 M2} + \gamma_{M1 M1} + \gamma_{E1 E1} \right).
\eea

We notice from Eqs.~(\ref{lex1})--(\ref{lex2}) and Eqs.~(\ref{b1c0})--(\ref{b2c0}) that the electric and magnetic dipole polarizabilities measured in RCS fully determine the terms of order $\nu^2$ and $Q^2$ in the LEXs of both VVCS amplitudes $T_1$ and $T_2$. 
In order to fully specify the LEXs for the spin-dependent forward VCS amplitudes $S_1$, and $S_2$, we need in addition the coefficients $b_{8, 0}$ and  $b_{21, 0}$.
We next show how they can be related to two of the generalized polarizabilities (GPs), determined
from the (non-forward) VCS process 
\bea
\gamma^*(q) + N(p) \to \gamma(q^\prime) + N(p^\prime),
\label{eq:vcs} 
\eea
where the outgoing photon is real and carries a low momentum, i.e. $q^{\prime \, 2} = 0$ and $q^\prime \to 0$.  

The VCS experiments at low outgoing photon energies 
can also be analyzed in terms of LEXs, as proposed in Ref.~\cite{Guichon:1995pu}. The VCS tensor describing the process (\ref{eq:vcs})  has been split in Ref.~\cite{Guichon:1995pu} into a Born part, which is defined as the nucleon intermediate state contribution using the $\gamma^\ast \gamma N$ vertex of Eq.~(\ref{emvertex}), and a non-Born part. The latter describes the response of the nucleon to the quasi-static electromagnetic field, due to the nucleon's internal structure. For the lowest-order nucleon-structure terms, one considers the response linear in the energy of the produced real photon.  
The VCS tensor describing the process (\ref{eq:vcs}) can generally be parametrized in terms of 12 independent amplitudes. In Ref.~\cite{DKK98}, a gauge-invariant tensor basis was constructed such that the non-Born invariant amplitudes are free of kinematical singularities and constraints:
\bea
M^{\mu \nu} &=& \sum_{i = 1}^{12} f_i(q^2, q \cdot q^\prime, q \cdot P) \, \rho_i^{\mu \nu} ,
\label{eq:vcstensor}
\eea
where the explicit expression for the tensors $\rho_i^{\mu \nu}$  can be found in 
Ref.~\cite{DKK98}. Furthermore in the limit $q^{\prime \, 2} = 0$, 
the 12 invariant amplitudes $f_i$ are related with the invariants $B_i$ 
of Eq.~(\ref{eq:dvcs}), describing the doubly-virtual Compton scattering process:
\bea
f_1 = B_1, \quad \quad && f_2 = B_2, \hspace{2.6cm} f_3 = B_4,  \nonumber \\
f_4 = B_7, \quad \quad && f_5 = B_8 - B_9, \hspace{1.75cm} f_6 = B_{10},  \nonumber \\
f_7 = B_{11}, \quad \quad && f_8 = B_{12} + B_{13}, \hspace{1.5cm} f_9 = B_{14},  \nonumber \\
f_{10} = B_{17}, \quad \quad && f_{11} = B_{18}, \hspace{2.25cm} f_{12} = B_{20} + B_{21},
\label{eq:btof}
\eea 
where the limit $q^{\prime \, 2} = 0$ is taken in the argument of the $B_i$. 

The behavior of the non-Born VCS tensor
at low energy ($q^\prime \to 0$)
but at arbitrary three-momentum $\bar {\rm q}$ of the virtual photon, which is conveniently defined in the c.m.\ system of the $\gamma^\ast  N$ system, can be parametrized by six independent GPs~\cite{Guichon:1995pu,DKK98}. The GPs can be accessed in experiment through the 
$e N \to e N \gamma$ process; see the reviews \cite{Guichon:1998xv,Drechsel:2002ar} for more details. 
At lowest order in the outgoing photon energy, there are two spin-independent GPs, denoted by 
$P^{(L1, L1)0}$, $P^{(M1, M1)0}$, 
and four spin GPs, denoted by $P^{(L1, M2)1}$, $P^{(M1, L2)1}$, $P^{(L1, L1)1}$, and $P^{(M1, M1)1}$, 
which are all functions of $Q^2$.\footnote{Equivalently, they can be considered
as functions of $\bar{\rm q}^2=Q^2(1+ \tau)$; this definition
is used in Ref.~\cite{Guichon:1995pu}.}
In this notation, $L$ stands for the longitudinal (or electric) and $M$ for the magnetic nature of the transition respectively.   
At $Q^2 = 0$, four of the six GPs are related to the polarizabilities from RCS as
\bea
\alpha_E &=& - \alpha_\mathrm{em} \frac{\sqrt{3}}{\sqrt{2}} \, P^{(L1, L1)0}(0), \hspace{2cm}
\beta_M = - \alpha_\mathrm{em} \frac{\sqrt{3}}{2 \sqrt{2}} \, P^{(M1, M1)0}(0), \nonumber \\
\gamma_{E1 M2} &=& - \alpha_\mathrm{em} \frac{3}{\sqrt{2}} \, P^{(L1, M2)1}(0), \hspace{1.2cm}
\gamma_{M1 E2} = - \alpha_\mathrm{em} \frac{3 \sqrt{3}}{2 \sqrt{2}} \, P^{(M1, L2)1}(0), 
\eea
whereas the remaining two GPs vanish in the real photon limit, i.e. $P^{(L1, L1)1} (0) = 0$, and $P^{(M1, M1)1} (0) = 0$. 

The GPs can be expressed through the non-Born ($\mathrm{nB}$) parts of the invariant amplitudes $f_i$. 
Using the shorthand notation,
\bea
\bar f_i(Q^2) \equiv f_i^\mathrm{nB}(q^2 = -Q^2, 0, 0), 
\eea
together with Eq.~(\ref{eq:btof}), 
these expressions are~\cite{DKK98}:
\bea
b_{8, 0}^\mathrm{nB} &=& \bar f_5(0) = - 6 M  P^{\prime \, (M1,M1)1}(0), 
\label{b8c0} \\
b_{21, 0}^\mathrm{nB} &=& \bar f_{12}(0) = \frac{3}{2}  
\left[ P^{\prime (M1, M1)1}(0) -  P^{\prime \, (L1, L1)1} (0) \right]
+ \frac{1}{\alpha_\mathrm{em}} \frac{1}{2 M} \, \gamma_{M1E2}. 
\label{b21c0}
\eea
In Eqs.~(\ref{b8c0}, \ref{b21c0}) we have introduced the notations for the slopes at $Q^2 =0$ of two GPs as:
\bea
\eqlab{GPslope}
P^{\prime\,  (L1, L1)1}(0) &\equiv& \frac{\rm d\hphantom{Q^2}}{{\rm{d}} Q^2} 
 P^{ (L1, L1)1}(Q^2 )  \bigg|_{Q^2 =0}, \\
P^{\prime\,  (M1, M1)1}(0) &\equiv& \frac{\rm d\hphantom{Q^2}}{{\rm d} Q^2} 
 P^{ (M1, M1)1}(Q^2 )  \bigg|_{Q^2 =0}. 
 \eea

We note that the lowest-order polarizabilities as measured through RCS together with the slopes at $Q^2 = 0$ of the two lowest-order GPs which themselves vanish at $Q^2 = 0$, and thus require a measurement through the VCS process, specify all low-energy constants appearing in the VVCS amplitudes of Eqs.~(\ref{lex1}-\ref{lex4}). 

\subsection{VVCS sum rules}
\label{sec4}

Having established the LEXs of the forward double VCS amplitudes $T_1, T_2, S_1$ and $S_2$,
we are ready to use the analyticity in  $\nu$,  for fixed spacelike photon virtuality, i.e. $Q^2 \geq 0$.
We distinguish two cases depending on their symmetry under $s \leftrightarrow u$ crossing, which flips the sign of  $\nu$: the amplitudes $T_1, T_2$ and $S_1$ are even functions of $\nu$ whereas $S_2$ is 
odd. We will present the relations for the non-pole parts of the amplitudes,  
$T_1^\mathrm{np}(\nu,\,Q^2)= T_1(\nu,\,Q^2) - T_1^\mathrm{pole}(\nu,\,Q^2)$  etc., i.e., the
well-known pole amplitudes given by Eq.~(\ref{pole}) are subtracted from the full amplitudes. 

\subsubsection{Spin-independent amplitude $T_1$}

The dispersion relation for $T_1$ requires one subtraction, which we take at $\nu = 0$, in order to ensure high-energy convergence~: 
\bea
\re T_1^\mathrm{np}(\nu,\,Q^2)\, & = & \,
\re T_1^\mathrm{np}(0,\,Q^2) + \frac{\nu^2}{2\pi}\,{\mathcal{P}}\,
\int_{\nu_0}^{\infty}\, d\nu^\prime \, \frac {1}{\nu^\prime (\nu^{\prime \,2} -\nu^2)}\, \frac{e^2}{M} F_1(x^\prime, Q^2),
\label{eq:T1dr} 
\eea
with $x^\prime \equiv Q^2 / (2 M \nu^\prime)$. 
Because the non-pole amplitudes are analytic
functions of $\nu$, they can be expanded in a Taylor series about $\nu=0$ with
a convergence radius determined by the lowest singularity, the threshold of pion production
at $\nu=\nu_0$. Analogous to the low-energy expansion of RCS,
the series in $\nu$, at fixed value of $Q^2$, for  forward double VCS takes the following form~\cite{Drechsel:2002ar}:
\bea
 T_1^\mathrm{np}(\nu,\,Q^2) \, & = &  T_1^\mathrm{np}(0,\,Q^2)
\,+\, \left( \alpha_E(Q^2) + \beta_M(Q^2) \right) \, \nu^2 \,+\,{\mathcal{O}}(\nu^4) \,,
\label{eq:T1lex}
\eea
The coefficients of the Taylor series of Eq.~(\ref{eq:T1lex}) follow by expanding the dispersion integrals 
as function of $\nu$. This yields a generalization of Baldin's sum rule for the forward dipole polarizabilities~\cite{Drechsel:2002ar}:
\bea
\label{eq:generalbaldin}
\alpha_E(Q^2) + \beta_M(Q^2) & \;=\; & \frac{e^2 M}{\pi \, Q^4}\,\int_{0}^{x_0} dx \, 
2x\,F_1(x,\,Q^2)\, ,
\eea
where $x_0$ corresponds with the pion production threshold. 
We next discuss the subtraction function at $\nu = 0$, $T_1^\mathrm{np}(0,\,Q^2)$,  
entering the dispersion relation of Eq.~(\ref{eq:T1dr}). 
Although in general the $Q^2$ behavior of this function is unknown, one can 
express its behavior at low $Q^2$ in terms of polarizabilities, see, e.g., Ref.~\cite{Bernabeu:1976jq}.  
We like to emphasize that polarizabilities are conventionally defined by separating 
the Compton amplitudes into Born and non-Born parts, with Born parts given by Eqs.~(\ref{born}).
The non-Born part  of  $T_1$ can then be read off Eqs.~(\ref{lex1}), (\ref{b1c0}), (\ref{b2c0}) as
\begin{eqnarray}
T_1^\mathrm{nB}(\nu,Q^2) 
&=&  \left( \alpha_E +\beta_M \right) \nu^2 
		+ \beta_M	Q^2\,  + \mathcal{O}(k^4), 
\label{eq:T1nb}
\end{eqnarray}
with $k = \{\nu, Q \}$.  
To obtain the low-energy expansion in $k$ of the non-pole part $T_1^\mathrm{np}$ 
entering Eq.~(\ref{eq:T1dr}), 
we also need to account for the difference between the Born and pole parts, which can be easily read off 
Eq.~(\ref{born}) as
\begin{eqnarray}
T_1^\mathrm{Born}(\nu,Q^2) - T_1^\mathrm{pole}(\nu,Q^2) &=& - \frac{\alpha_\mathrm{em}}{M} F_D^2 = 
- \frac{\alpha_\mathrm{em} }{M} e_N^2 + \frac{\alpha_\mathrm{em} }{3 M} e_N \langle r_1^2 \rangle \, Q^2 + 
\mathcal{O}(Q^4)\, , 
\label{eq:T1bmp}
\end{eqnarray}
where $\langle r_1^2 \rangle$ is the squared Dirac radius of the nucleon. 
Combining Eqs.~(\ref{eq:T1nb}) and (\ref{eq:T1bmp}), one then obtains the low-energy expansion 
of $T_1^{np}$  in both $\nu^2$ and $Q^2$ as
\begin{eqnarray}
T_1^\mathrm{np}(\nu,Q^2) 
&=&  - \frac{\alpha_\mathrm{em} }{M} e_N^2 + \left( \alpha_E +\beta_M \right) \nu^2 
	+ \left(\frac{\alpha_\mathrm{em} }{3 M} e_N \langle r_1^2 \rangle + \beta_M	\right) Q^2\,  + 
		\mathcal{O}(k^4).
\label{eq:T1np}
\end{eqnarray}
Consequently, the subtraction function at $\nu = 0$, which enters the dispersion relations of 
Eq.~(\ref{eq:T1dr}), is given up to terms of order $\mathcal{O}(Q^4)$ by
 \begin{eqnarray}
T_1^\mathrm{np}(0,Q^2) 
&=&  - \frac{\alpha_\mathrm{em} }{M} e_N^2 + \left(\frac{\alpha_\mathrm{em} }{3 M} e_N \langle r_1^2 \rangle + \beta_M \right) Q^2\,  + \mathcal{O}(Q^4). 
\label{eq:T1np0}
\end{eqnarray}

\subsubsection{Spin-independent amplitude $T_2$}

For the amplitude $T_2$, which is even in $\nu$,  one can write down an unsubtracted DR in $\nu$:
\bea
\re T_2^\mathrm{np}(\nu,\,Q^2)\, =  \,
 \frac{1}{2 \pi} \,{\mathcal{P}}\,
 \int_{\nu_0}^{\infty}\, d\nu^\prime \,   \frac{1}{\nu^{\prime \, 2} - \nu^2}  \, e^2 F_2(x^\prime ,\,Q^2)\,.
\label{eq:T2dr}
\eea
The expansion of the amplitude $T_2$ at small $k = \{\nu, Q\}$ can 
be read off Eqs.~(\ref{lex1}, \ref{b2c0}) as
\footnote{For the amplitude $T_2$, there is no difference between the Born and pole contributions, as seen from Eq.~(\ref{born}).}
\bea
T_2^\mathrm{np}(\nu,Q^2) 
=  \left( \alpha_E +\beta_M \right) Q^2 + \mathcal{O}(k^4).
\label{eq:T2nplex}
\eea
By evaluating Eq.~(\ref{eq:T2dr}) at $\nu= 0$, taking its derivative with respect to $Q^2$ at $Q^2 = 0$,  
and using the relation
\bea
\left[ \frac{1}{Q^2} F_2 (x, Q^2) \right]_{Q^2 = 0} = \left[ \frac{1}{Q^2} 2 x F_1 (x, Q^2) \right]_{Q^2 = 0} = 
\frac{1}{e^2 \, \pi} \sigma_T,
\eea 
with $\sigma_T$ the total (real) photon absorption cross section, 
one recovers the Baldin sum rule~\cite{Baldin}, i.e. Eq.~(\ref{eq:generalbaldin}) evaluated at $Q^2 = 0$, 
for $(\alpha_E + \beta_M)$.

\subsubsection{Spin-dependent amplitude $S_1$}

We next discuss the DR for the spin-dependent
amplitude $S_1$. The amplitude $S_1$ is
even in $\nu$, and the unsubtracted DR for its non-pole part reads
\be
\label{eq:S1dr}
\re S_1^\mathrm{np}(\nu,\,Q^2) \,=\,
\frac{1}{2 \pi}\,{\mathcal{P}}\,\int_{\nu_0}^{\infty} \, d\nu' \, 
\frac{\nu^\prime}{\nu^{\prime \, 2} - \nu^2} \, \frac{e^2}{\nu^\prime} \, g_1(x^\prime,\,Q^2).
\ee
The low-energy expansion in $\nu$, at fixed value of $Q^2$, for $S_1^{np}$ takes the form~\cite{Drechsel:2002ar}
\be
\label{S1lexnu}
S_1^\mathrm{np}(\nu,\,Q^2) =
\frac{2 \, \alpha_\mathrm{em}}{M } \,  I_1(Q^2)
+ \left[ \frac{2 \, \alpha_\mathrm{em} }{M\, Q^2}
\bigg( I_{TT}(Q^2) - I_1(Q^2) \bigg)
+ M \delta_{LT}(Q^2) \right] \, \nu^2  \,+\, {\mathcal{O}}(\nu^4),
\ee
where the leading term of ${\mathcal {O}(\nu^0)}$ follows
from Eq.~(\ref{eq:S1dr}) as
\bea
\label{I1sr}
I_1(Q^2) &\; = \;& \frac{2M^2}{Q^2}\int_0^{x_0}
dx \, g_1(x,\,Q^2)  . 
\eea
Using Eqs.~(\ref{born}) and (\ref{lex3}) one obtains the low-energy theorem result~: 
$S_1^{np}(0,\, 0) = - \alpha_{em} \kappa_N^2 / (2 M)$, which yields the GDH
sum rule for real photons~\cite{Gerasimov:1965et,Drell:1966jv}, 
$I_1(0) = - \kappa_N^2 / 4$. The $\nu^2$-dependent term in the expansion of Eq.~(\ref{S1lexnu}) involves, besides $I_1$, 
also the moment $I_{TT}$ of the helicity difference cross sections and a longitudinal-transverse polarizability $\delta_{LT}$, which are expressed through moments of spin structure functions as~\cite{Drechsel:2002ar}
\bea
I_{TT}(Q^2) &\;=\;& \frac{2 M^2}{Q^2}\,
\int_{0}^{x_0}\,dx \, \left\{ g_1\,(x,\,Q^2)
\,-\, \frac{4 M^2}{Q^2} \, x^2 \, g_2\,(x,\,Q^2) \right\} , \label{ITT}  \\
\delta_{LT}\,(Q^2) 
&\;=\;& \frac{4 e^2 M^2}{\pi Q^6}\,\int_{0}^{x_0}\,dx \, x^2 \,
\left\{ g_1\,(x,\,Q^2) \,+\, g_2\,(x,\,Q^2) \right\} .
\label{deltaLT}
\eea
At $Q^2 = 0$, the $\nu^2$ term in the low-energy expansion of $S_1^\mathrm{np}$
can be read off Eqs.~(\ref{lex3}) and (\ref{b7c1}), yielding~\footnote{Note that this implies the relation~\cite{Drechsel:2002ar}: $I_{TT}^\prime (0) -  I_1^\prime (0) = \frac{M^2}{2 \, \alpha_\mathrm{em}} \, \left(\gamma_0 - \delta_{LT} \right) $, with $I_i^\prime(0) \equiv \frac{d}{d Q^2} I_i(Q^2) \bigg|_{Q^2 = 0}$, and $\delta_{LT} \equiv \delta_{LT}(0)$. \label{footnote2}}
\be
\label{S1lexnuq0}
S_1^\mathrm{np}(\nu,\, 0) =
\frac{2 \, \alpha_\mathrm{em}}{M } \,  \left( - \frac{\kappa_N^2}{4} \right)
+   M \, \gamma_0 \,  \nu^2  \,+\, {\mathcal{O}}(\nu^4)\,,
\ee
where $\gamma_0$ is the forward spin polarizability as accessed in RCS, which can be obtained as the $Q^2 \to 0$ 
limit of the integral obtained in Ref.~\cite{Drechsel:2002ar}: 
\bea
\label{gamma0sr}
\gamma_0\,(Q^2) &\;=\;& \frac{ 4 M^2 e^2 \,}{\pi  Q^6}\,\int_{0}^{x_0}\,dx \, x^2 \,
\left\{ g_1\,(x,\,Q^2)
\,-\, \frac{4 M^2}{Q^2} \, x^2 \, g_2\,(x,\,Q^2) \right\} .
\eea

We can derive a new sum rule by performing a Taylor series in $Q^2$ at $\nu = 0$ for $S_1^\mathrm{np}$. By expanding $I_1(Q^2)$ in Eq.~(\ref{S1lexnu}), we obtain
\bea
\label{S1lexqsqr}
S_1^\mathrm{np}(0,\,Q^2) =
\frac{2 \, \alpha_\mathrm{em}}{M } \,  \left\{ - \frac{\kappa_N^2}{4} + Q^2 \, I^\prime_1(0)  + {\mathcal{O}}(Q^4) \right\}, 
\eea
where $I_1^\prime(0) \equiv \frac{d}{d Q^2} I_1(Q^2) \bigg|_{Q^2 = 0}$ is the $Q^2$ slope at $Q^2 = 0$ of the first moment of the structure function $g_1$.
Using the low-energy expansion of Eq.~(\ref{lex3}), we can identify the $Q^2$-dependent term of the non-Born part 
$S_1^\mathrm{nB}$ at $\nu = 0$ as
\bea
S_1^\mathrm{nB}(0, Q^2) &=& \alpha_\mathrm{em} M Q^2 \left\{ 4 M \, b_{10,0}^\mathrm{nB} + 4 \, b_{18,0}^\mathrm{nB}  + b_{8,0}^\mathrm{nB} + 2 M \, b_{21,0}^\mathrm{nB} \right\} + {\cal O}(Q^4),  \nonumber \\
&=& 
\alpha_\mathrm{em} M Q^2 \left\{ \frac{1}{\alpha_\mathrm{em}} \gamma_{E1 M2} 
- 3 M \left[ P^{\prime \, (M1, M1)1}(0) +  P^{\prime (L1, L1)1} (0) \right]   \right\} + {\cal O}(Q^4),  
\label{S1nb}
\eea
where in the last line we have used Eqs.~(\ref{b10c0}), (\ref{b18c0}), (\ref{b8c0}), (\ref{b21c0}) 
for the corresponding low-energy coefficients. 
To relate $I^\prime_1(0)$ with the expression in Eq.~(\ref{S1nb}), we need to account for the difference between 
Born and pole parts, which can be read off Eq.~(\ref{born}) as
\begin{eqnarray}
S_1^\mathrm{Born}(\nu,Q^2) - S_1^\mathrm{pole}(\nu,Q^2) &=& - \frac{\alpha_\mathrm{em}}{2 M} F_P^2 = 
\frac{2 \alpha_\mathrm{em} }{M} \left\{ - \frac{\kappa_N^2}{4} + \frac{ \kappa_N^2}{12} \langle r_2^2 \rangle \, Q^2 + \mathcal{O}(Q^4) \right\} \, , 
\label{S1bmp}
\end{eqnarray}
where $\langle r_2^2 \rangle$ is the nucleon mean squared Pauli radius. 
Combining Eqs.~(\ref{S1lexqsqr}), (\ref{S1nb}), and (\ref{S1bmp}) then allows us to derive a new sum rule 
relating the slope at $Q^2 = 0$ of the GDH sum rule to the Pauli radius and polarizabilities as measured in RCS and VCS:
\bea
I_1^\prime(0) =  \frac{ \kappa_N^2}{12} \langle r_2^2 \rangle 
+ \frac{M^2}{2} \left\{ \frac{1}{\alpha_\mathrm{em}} \gamma_{E1 M2} 
- 3 M \left[ P^{\prime \, (M1, M1)1}(0) +  P^{\prime \, (L1, L1)1} (0) \right]  \right\}. 
\label{qsqrgdhsr}
\eea
We like to emphasize that all quantities entering Eq.~(\ref{qsqrgdhsr}) are observable quantities:  
the {\it lhs} is obtained from the first moment of the spin structure function $g_1$~\cite{Kuhn:2008sy,Chen:2010qc}, 
whereas the {\it rhs} involves the Pauli radius as well as spin polarizabilities measured through the RCS and VCS processes. Therefore the sum rule of Eq.~(\ref{qsqrgdhsr}) provides us with a model-independent and predictive relation.  In the next sections, we will test this new GDH sum rule for finite photon virtuality using heavy-baryon as well as covariant baryon chiral perturbation theory. We will also provide a phenomenological evaluation based on available data.

\subsubsection{Spin-dependent amplitude $S_2$}

Finally, for the second spin-dependent forward double VCS amplitude $S_2$, which is odd in $\nu$, an 
unsubtracted DR takes the form
\be
\label{S2dr}
\re S_2(\nu,\,Q^2)
\,=\,\re S_2^\mathrm{pole}(\nu,\,Q^2) \, + \, \frac{\nu}{2 \pi} \,{\mathcal{P}}\,
\int_{\nu_0}^{\infty} d\nu^\prime  \frac{1}{\nu^{\prime \, 2} -\nu^2} \, \frac{e^2 M}{\nu^{\prime \, 2}} \, g_2(x^\prime,\,Q^2) \, .
\ee
If we further assume that the amplitude $S_2$ converges faster than $1/\nu$ for
$\nu \rightarrow \infty$,
we may write an unsubtracted dispersion relation for
the amplitude $\nu \, S_2$, which is even in $\nu$,
\be
\label{S2dr2}
\re \left[\nu \, S_2(\nu,\,Q^2)\right]
\,=\,  \re \left[\nu \, S_2(\nu,\,Q^2)\right]^\mathrm{pole}
\,+\, \frac{1}{2 \pi} \,{\mathcal{P}} \,  \int_{\nu_0}^{\infty} \,d\nu^\prime \,
\frac{1}{\nu^{\prime \, 2} -\nu^2} \, e^2 M \, g_2(x^\prime,\,Q^2) \, .
\ee
If we now multiply Eq.~(\ref{S2dr}) by $\nu$ and subtract it from Eq.~(\ref{S2dr2}), we obtain the  Burkhardt-Cottingham (BC) ``superconvergence sum rule''~\cite{Burkhardt:1970ti},
valid for {\it{any}} value of $Q^2$:
\be
\label{BC}
\int_{0}^{1}\, g_{2}\,(x,\,Q^2)\, dx  = 0 ,
\ee
provided that the integral converges for $x \rightarrow 0$. 
Notice that the upper integration limit in the integral of Eq.~(\ref{BC}) extends to $1$, and thus includes the elastic contribution. 
By separating the elastic and inelastic parts in the integral of Eq.~(\ref{BC}), the BC sum rule can be cast into the 
equivalent form
\bea
\label{BC2}
I_2(Q^2)& \equiv & \frac{2M^2}{Q^2}\int_0^{x_0}g_2(x,\,Q^2)\,dx \,=\,
\frac{1}{4} \, F_P(Q^2) G_M(Q^2)  .
\eea
The BC sum rule was shown to be satisfied in quantum
electrodynamics by an explicit calculation at lowest order in the coupling constant
$\alpha_\mathrm{em}$ \cite{Tsa75}. In perturbative QCD, the BC sum rule
was verified for a quark target to first order in $\alpha_s$~\cite{Alt94}. 
Furthermore, in the non-perturbative domain of low $Q^2$, the BC sum rule was also verified within heavy-baryon chiral perturbation theory \cite{Kao:2002cp, Kao:2003jd}. 

The LEX of $(\nu S_2)^\mathrm{np}$ can be expressed as~\cite{Drechsel:2002ar}
\bea
\label{S2lexnu}
\left[\nu \, S_2(\nu,\,Q^2)\right]^\mathrm{np} =
2 \, \alpha_\mathrm{em} \, I_2(Q^2)
\,+\, \frac{2 \, \alpha_\mathrm{em}}{Q^2} \, I_2^{(3)}(Q^2)  \, \nu^2
\,+\, {\mathcal{O}}(\nu^4) \,,
\eea
where the observable $I_2^{(3)}(Q^2)$ is defined 
through the third moment of the spin structure function $g_2$ as
\bea
I_2^{(3)}(Q^2) &\equiv&  \frac{8 M^4}{Q^4}\, \int_{0}^{x_0}\,dx \,  x^2 \, g_2\,(x,\,Q^2),  
\label{g2thirdmom} 
\\
&=& I_{1}(Q^2) - I_{TT}(Q^2), 
\eea
and where the last line has been obtained by using Eqs.~(\ref{I1sr}) and (\ref{ITT}). 
Note that the slope at $Q^2 = 0$ of  $I_2^{(3)}(Q^2) $ follows from footnote~\ref{footnote2} as
\bea
 I_2^{(3) \, \prime}(0) &=& \frac{M^2}{2 \alpha_{em}} \left( \delta_{LT} - \gamma_0 \right). 
 \label{eq:slopeI23}
\eea
Using the low-energy expansion of Eq.~(\ref{lex4}), we can identify the $\nu^2$ dependent term of the non-Born part 
of $\nu S_2^\mathrm{nB}$ as
\bea
\nu S_2^\mathrm{nB}(\nu, Q^2) &=& \alpha_\mathrm{em} M^3 \nu^2 \left\{ 4 \, b_{10,0}^\mathrm{nB} - 4 \, b_{17,1}^\mathrm{nB} + 2 \, b_{21,0}^\mathrm{nB} \right\} + {\cal O}(\nu^4, \nu^2 Q^2),  
\label{S2nb}
\eea
To relate Eqs.~(\ref{S2lexnu}) and (\ref{S2nb}), we need to account for the difference between 
Born and pole parts, which can be read off Eq.~(\ref{born}) as
\begin{eqnarray}
\nu S_2^\mathrm{Born}(\nu,Q^2) - \left[ \nu S_2 (\nu,Q^2) \right]^\mathrm{pole} &=& \frac{\alpha_\mathrm{em}}{2} F_P(Q^2) G_M(Q^2) ,\label{S2bmp}
\end{eqnarray}
and precisely accounts for the leading term of ${\mathcal {O}(\nu^0)}$ in Eq.~(\ref{S2lexnu}), as given by the BC sum rule, Eq.~(\ref{BC2}).
The terms of ${\mathcal {O}(\nu^2)}$ in Eqs.~(\ref{S2lexnu}) and (\ref{S2nb}) can then be identified to yield the 
new sum rule:
\bea
I_2^{(3) \, \prime}(0) &=& M^3 \left\{ 2 \, b_{10,0}^\mathrm{nB} - 2 \, b_{17,1}^\mathrm{nB} + b_{21,0}^\mathrm{nB} \right\} .  
\label{S2sr1}
\eea
On the {\it rhs} of Eq.~(\ref{S2sr1}), the low-energy quantities $b_{10,0}^\mathrm{nB}$, $b_{17,1}^\mathrm{nB}$,  and $b_{21,0}^\mathrm{nB}$ are related to polarizabilities as measured in RCS and VCS through Eqs.~(\ref{b10c0}), (\ref{b17c1}), and (\ref{b21c0}). In this way we obtain the following sum rule:
\bea
I_2^{(3) \, \prime}(0) &=& \frac{M^2}{2} 
\left\{- \frac{1}{\alpha_\mathrm{em}} \left[ \gamma_0 + \gamma_{E1 E1} \right]
+ 3 M \left[ P^{\prime \, (M1, M1)1}(0) -  P^{\prime \, (L1, L1)1} (0) \right]   \right\}. 
\label{S2sr2}
\eea 
Using Eq.~(\ref{eq:slopeI23}), the sum rule of Eq.~(\ref{S2sr2}) can be expressed equivalently as
\bea
\delta_{LT} &=& 
- \gamma_{E1 E1} 
+ 3 M \alpha_\mathrm{em} \, \left[ P^{\prime \, (M1, M1)1}(0) -  P^{\prime \, (L1, L1)1} (0) \right]. 
\label{S2sr3}
\eea
Note that similar to its counterpart of Eq.~(\ref{qsqrgdhsr}), all quantities which enter Eq.~(\ref{S2sr3}) are observables. Therefore the new sum rule of Eq.~(\ref{S2sr3}) provides us with a second model-independent and predictive relation among low-energy spin structure constants of the nucleon.

\section{Verification in heavy-baryon chiral perturbation theory}
\label{sec5}

In this section we verify the new GDH sum rule of Eq.~(\ref{qsqrgdhsr}) for finite photon virtuality as 
well as the sum rule of Eq.~(\ref{S2sr3}) for $\delta_{LT}$ within the context 
of heavy-baryon chiral perturbation theory (HB$\chi$PT). 
For the purpose of this verification, we will express the two sum rules of 
Eqs.~(\ref{qsqrgdhsr}) and (\ref{S2sr3}) equivalently as relations for the GPs as
\bea
P^{\prime \, (M1, M1)1}(0)  &=& 
\frac{1}{6 M} \left\{ \frac{2}{M^2} \left( \frac{ \kappa_N^2}{12} \langle r_2^2 \rangle - I_1^\prime(0) \right) 
+ \frac{1}{\alpha_\mathrm{em}} \left( \gamma_{E1 M2} + \gamma_{E1 E1}  + \delta_{LT}  \right)
\right\}, 
\label{srm1m1}\\
P^{\prime \, (L1, L1)1}(0)  &=& 
\frac{1}{6 M} \left\{ \frac{2}{M^2} \left( \frac{ \kappa_N^2}{12} \langle r_2^2 \rangle - I_1^\prime(0) \right) 
+ \frac{1}{\alpha_\mathrm{em}} \left( \gamma_{E1 M2} - \gamma_{E1 E1}  - \delta_{LT}  \right)
\right\}.
\label{srl1l1}
\eea

In HB$\chi$PT, the leading order (LO) contribution in the chiral expansion of both of these sum rules corresponds with terms which are proportional to $1/m_\pi^2$, with $m_\pi$ the pion mass. The next-to-leading order (NLO) contribution corresponds with terms proportional to $1/(m_\pi M)$. 
All terms which are necessary to verify this sum rule have been calculated already up to NLO in the literature. 

The evaluation of Eqs.~(\ref{srm1m1}), (\ref{srl1l1}) involves the first moment $I_1(Q^2)$ of the $g_1$ structure function, which was evaluated in HB$\chi$PT up to NLO in the chiral expansion as~\cite{Ji:1999pd, Ji:1999mr}
\begin{eqnarray}
I_1^\prime(0) &=& \frac{g_{A}^{2}}{(4\pi f_{\pi})^2} \, \frac{M}{m_\pi} \, \frac{\pi}{48}
\left[ 1 + 3 \kappa_V + (2 + 6 \kappa_S) \tau_{3} \right],
\label{I1chpt}
\end{eqnarray}
where $f_\pi = 92.4$ MeV is the pion decay constant, $g_A = 1.27$ is the nucleon axial coupling constant, 
$\tau_3$ is the third Pauli isospin matrix, and $\kappa_V = 3.70$ ($\kappa_S = -0.12$) denotes the nucleon isovector (isoscalar) anomalous magnetic moment. 

Furthermore, the HB$\chi$PT calculation of $\delta_{LT}$, which appears 
in both Eqs.~(\ref{srm1m1}) and (\ref{srl1l1}), was performed in Ref.~\cite{Kao:2002cp} up to NLO in the chiral expansion: 
\begin{eqnarray}
\frac{1}{\alpha_\mathrm{em}} \, \delta_{LT} &=& \frac{g_{A}^{2}}{(4\pi f_{\pi})^2} \, \frac{1}{m_\pi^2} \, \frac{1}{3}
\left\{ 1 + \frac{\pi}{8} \, \frac{m_\pi}{M}
\left[ -3 + \kappa_V + (-6 + 4 \kappa_S) \tau_{3} \right]
\right\}. 
\label{deltaltchpt}
\end{eqnarray}

The first term on the {\it rhs} of Eqs.~(\ref{srm1m1})--(\ref{srl1l1}) involves the Pauli radius. To the order needed, its expression in HB$\chi$PT is given by~\cite{BKM95}
\begin{eqnarray}
\frac{\kappa_N^2}{12} \, \langle r_2^2 \rangle = - \frac{\kappa_N}{2} \, F_P^\prime(0) 
= \frac{g_{A}^{2}}{(4\pi f_{\pi})^2} \, \frac{M}{m_\pi} \, \frac{\pi}{24}
\left[ \kappa_V + \kappa_S \tau_{3} \right].
\label{r2chpt}
\end{eqnarray}

The nucleon spin polarizabilities $\gamma_{E1, M2}$ and $\gamma_{E1, E1}$, obtained from RCS, which appear on the {\it rhs} of Eqs.~(\ref{srm1m1})--(\ref{srl1l1}), are given up to NLO in the chiral expansion, i.e. ${\mathcal O}(p^4)$,  as~\cite{VijayaKumar:2000pv}
\begin{eqnarray}
\frac{1}{\alpha_\mathrm{em}} \gamma_{E1 M2} &=& \frac{g_{A}^{2}}{(4\pi f_{\pi})^2} \, \frac{1}{m_\pi^2} \, \frac{1}{6} 
\left\{ 1 - \frac{\pi}{4} \, \frac{m_\pi}{M} \left[ 6 + \tau_{3} \right] \right\}, \\
\frac{1}{\alpha_\mathrm{em}} \gamma_{E1 E1} &=& \frac{g_{A}^{2}}{(4\pi f_{\pi})^2} \, \frac{1}{m_\pi^2} \, \frac{1}{6} 
\left\{ -5 + \frac{\pi}{4} \, \frac{m_\pi}{M} \left[ 22 + 11 \, \tau_{3} \right] \right\}. 
\label{spinpolchpt}
\end{eqnarray}

In this way, we obtain for the {\it rhs} of Eqs.~(\ref{srm1m1})--(\ref{srl1l1})
\bea
\frac{1}{6 M} \left\{ \frac{2}{M^2} \left( \frac{ \kappa_N^2}{12} \langle r_2^2 \rangle - I_1^\prime(0) \right) 
+ \frac{1}{\alpha_\mathrm{em}} \left( \gamma_{E1 M2} + \gamma_{E1 E1}  + \delta_{LT}  \right)
\right\} 
&=& \frac{g_{A}^{2}}{(4\pi f_{\pi})^2}  \frac{1}{m_\pi^2} \frac{1}{18 M} 
\left\{ -1 + \frac{\pi}{4} \frac{m_\pi}{M} \left[ 6 + \tau_{3} \right] \right\}, \nonumber \\
\label{rhssrm1m1} \\
\frac{1}{6 M} \left\{ \frac{2}{M^2} \left( \frac{ \kappa_N^2}{12} \langle r_2^2 \rangle - I_1^\prime(0) \right) 
+ \frac{1}{\alpha_\mathrm{em}} \left( \gamma_{E1 M2} - \gamma_{E1 E1}  - \delta_{LT}  \right)
\right\} 
&=& \frac{g_{A}^{2}}{(4\pi f_{\pi})^2}  \frac{1}{m_\pi^2} \frac{1}{9 M}  \nonumber \\
&\times &
\left\{ 1 - \frac{\pi}{8} \frac{m_\pi}{M} \left[ 13 + \kappa_V + 4(1+ \kappa_S) \tau_{3} \right] \right\}. 
\label{rhssrl1l1}
\eea
To test these predictions in HB$\chi$PT, we need the derivatives of two GPs on the {\it lhs} of 
Eqs.~(\ref{srm1m1}, \ref{srl1l1}). They have been calculated 
at LO in the chiral expansion in Refs.~\cite{HHKS,Hemmert:1999pz} and at NLO in Refs.~\cite{Kao:2002cn, Kao:2004us}~\footnote{Note that some algebraic errors in the original version of Ref.~\cite{Kao:2004us} have been corrected in the corresponding erratum listed under Ref.~\cite{Kao:2004us}. We also note that for the GP $P^{(L1, L1)1}$ the NLO HB$\chi$PT result of Ref.~\cite{Kao:2004us} for the terms beyond the first derivative in $Q^2$ at $Q^2 = 0$ is incomplete as was pointed out by the recent covariant B$\chi$PT calculation~\cite{Lensky:2016nui}. In 
Ref.~\cite{Kao:2004us}, the NLO terms for the GP $P^{(L1, L1)1}$ were not calculated directly but inferred from nucleon crossing symmetry relations. As for the sum rule tests in the present work we only need the results for the first derivative $P^{\prime \, (L1, L1)1}(0)$; there is an exact agreement between the corresponding terms at both LO and NLO in the B$\chi$PT and HB$\chi$PT results.}. 
The derivatives at $Q^2 = 0$ appearing  in Eqs.~(\ref{srm1m1})--(\ref{srl1l1}) are given by
\bea
P^{\prime \, (M1, M1)1}(0) &=&  \frac{g_{A}^{2}}{(4\pi f_{\pi})^2} \, \frac{1}{m_\pi^2} \, \frac{1}{18 M}  
\left\{ -1 + \frac{\pi}{4} \, \frac{m_\pi}{M} \left[ 6 + \tau_{3} \right] \right\}, 
\label{m1m1chpt}\\
P^{\prime (L1, L1)1}(0) &=&  \frac{g_{A}^{2}}{(4\pi f_{\pi})^2} \, \frac{1}{m_\pi^2} \, \frac{1}{9 M}  
\left\{ 1 - \frac{\pi}{8} \, \frac{m_\pi}{M} \left[13 + \kappa_V+ 4( 1 + \kappa_S) \tau_{3} \right] \right\}.
\label{l1l1chpt}
\eea
We notice that both for the GP $P^{(M1,M1)1}$ of Eq.~(\ref{m1m1chpt})
and for the GP $P^{(L1,L1)1}$ of Eq.~(\ref{l1l1chpt})
there is an exact agreement both at LO and NLO with the 
{\it rhs} of the sum rule, Eqs.~(\ref{rhssrm1m1}) and~(\ref{rhssrl1l1}),
respectively.

We can also check the leading order contribution when including Delta states by
calculating the $\pi \Delta$ loop contributions in HB$\chi$PT using the small-scale expansion (SSE) to order {\cal O}$(\varepsilon^3)$. The leading contributions to the spin polarizabilities and to the GPs contain 
terms proportional to 
$1/{m_\pi}^2$ or $1/\varDelta^2$, where $\varDelta = M_\Delta - M_N$. Let us introduce the dimensionless ratios:
\be
\mu \equiv m_\pi/M, \quad \delta\equiv \varDelta/M.
\ee  
Then, denoting the leading $\pi N \Delta$ coupling constant by $h_A$, 
the Delta contributions to the various quantities which enter in the sum rules of 
Eqs.~(\ref{srm1m1}, \ref{srl1l1}) were calculated to order {\cal O}$(\varepsilon^3)$ 
in Refs.~\cite{Hemmert:1997tj,Hemmert:1999pz,Kao:2002cp,Kao:2003jd,Gockeler:2003ay}:
\begin{eqnarray}
P^{\prime \, (M1, M1)1}(0) \bigg|^{\pi\Delta}
&=& -  \frac{h_A^2}{1296 \pi^2 f_\pi^2 M M_\Delta^2} \, \frac{1}{\delta^2-\mu^2}
\left\{ 1 - \frac{\delta}{\sqrt{\delta^2 - \mu^2}}
\ln \frac{\delta + \sqrt{\delta^2 - \mu^2}}{\mu}  \right\} , 
 \label{sse1}\pagebreak[0] \\
P^{\prime \, (L1, L1)1}(0) \bigg|^{\pi\Delta}
&=& \frac{h_A^2}{648 \pi^2 f_\pi^2 M M_\Delta^2}  \, \frac{1}{\delta^2-\mu^2}
\left\{ 1 -\frac{\delta}{\sqrt{\delta^2 - \mu^2}}
\ln \frac{\delta + \sqrt{\delta^2 - \mu^2}}{\mu}  \right\} , 
\label{sse2}\pagebreak[0] \\
\frac{1}{\alpha_\mathrm{em}} \gamma_{E1, M2}\bigg|^{\pi\Delta} &=& 
\frac{h_A^2}{432 \pi^2 f_\pi^2 M_\Delta^2}  \, \frac{1}{\delta^2-\mu^2}
\left\{ 1 - \frac{\delta}{\sqrt{\delta^2 - \mu^2}}
\ln \frac{\delta + \sqrt{\delta^2 - \mu^2}}{\mu}  \right\},  
\label{sse3}\pagebreak[0] \\
\frac{1}{\alpha_\mathrm{em}} \gamma_{E1, E1}\bigg|^{\pi\Delta} &=& 
\frac{h_A^2}{432 \pi^2 f_\pi^2 M_\Delta^2}  \, \frac{1}{(\delta^2-\mu^2)^2}
\left\{\delta^2 + 5\mu^2+ \frac{(\delta^2 - 7\mu^2)\,\delta}{\sqrt{\delta^2 - \mu^2}}
\ln \frac{\delta + \sqrt{\delta^2 - \mu^2}}{\mu} \right\},  
\label{sse4}\pagebreak[0] \\
\frac{1}{\alpha_\mathrm{em}} \delta_{LT}\bigg|^{\pi\Delta} &=& 
\frac{h_A^2}{216 \pi^2 f_\pi^2 M_\Delta^2}  \, \frac{1}{(\delta^2-\mu^2)^2}
\left\{ -2 \delta^2 - \mu^2 + \frac{(\delta^2 + 2\mu^2) \, \delta}{\sqrt{\delta^2 - \mu^2}} \ln \frac{\delta + \sqrt{\delta^2 - \mu^2}}{\mu}   \right\},  
\label{sse5}\pagebreak[0] \\
I_1^\prime(0)\bigg|^{\pi\Delta} &=& 0,  
\label{sse6}\pagebreak[0] \\
\kappa_N^2 \, \langle r_2^2 \rangle \bigg|^{\pi\Delta} &=& 0.  
\label{sse7}
\end{eqnarray}
Plugging the leading order expressions of Eqs.~(\ref{sse1}-\ref{sse7}) into the sum rules of 
Eqs.~(\ref{srm1m1}, \ref{srl1l1}), one easily verifies that both sum rules are also exactly satisfied to 
this order in the SSE.

\section{Verification  in covariant $\chi$PT}
\label{sec6}

Using the results for the RCS spin polarizabilities and the VVCS amplitudes obtained in Refs.~\cite{Lensky:2009uv,Lensky:2014dda,Lensky:2014efa,Lensky:2015awa},
as well as the results for the nucleon GPs calculated in Ref.~\cite{Lensky:2016nui},
we have also verified the new sum rules in covariant B$\chi$PT.
Note that in practice it is more convenient to use the variant of the sum rules for the non-Born parts of the amplitudes $S_1$ and $S_2$, which read
\begin{eqnarray}
S_1^\mathrm{nB}(\nu,Q^2)&=&M\gamma_0\nu^2 + MQ^2\left\{\gamma_{E1M2}-3M\alpha_\mathrm{em} \left[ 
P^{\prime \, (M1,M1)1}(0)+P^{\prime \, (L1,L1)1}(0) \right] \right\}+\mathcal{O}(k^4), \\
\nu S_2^\mathrm{nB}(\nu,Q^2)&=&-M^2\nu^2\left\{
\gamma_0+\gamma_{E1E1}-3M\alpha_\mathrm{em} \left[ P^{\prime \, (M1,M1)1}(0) - P^{\prime \, (L1,L1)1}(0)
\right] \right\}+\mathcal{O}(\nu^4, \nu^2 Q^2)\,. 
\end{eqnarray}
For completeness, we provide here the results for the derivatives of the
two GPs in covariant B$\chi$PT that were obtained in Ref.~\cite{Lensky:2016nui}:
\begin{enumerate}
\item $\pi N$ loops:
\paragraph{Proton}
\begin{align}
P'^{(M1,M1)1}(0)|^{\pi N}_p & =\frac{g_A^2}{288\pi^2 f_\pi^2 M^3}\frac{1}{\mu^2(4-\mu^2)^{3/2}}
\nonumber\\
&\times \biggl[
\mu  \left(18 \mu ^8-143 \mu ^6+327 \mu ^4-192 \mu ^2+14\right) \arccos
\left(\frac{\mu ^2}{2}-1\right)\nonumber\\
&\hphantom{-}
+\sqrt{4-\mu ^2} \left(36 \mu ^6-160 \mu ^4+98 \mu ^2-4-\left(18 \mu ^6-107 \mu ^4+149 \mu ^2-36\right) \mu ^2 \ln
   \mu^2\right)\biggr]\nonumber\\
   &=-2.3~\mathrm{GeV}^{-5} \,,
\displaybreak[0]\label{eq:pmmpin}\\
P'^{(L1,L1)1}(0)|^{\pi N}_p & = \frac{g_A^2}{288\pi^2 f_\pi^2 M^3}\frac{1}{\mu^2(4-\mu^2)^{5/2}}
\nonumber\\
&\times
\biggl[
\mu  \left(27 \mu ^{10}-321 \mu ^8+1338 \mu ^6-2250
   \mu ^4+1288 \mu ^2-136\right) \arccos\left(\frac{\mu ^2}{2}-1\right)
   \nonumber\\
&\hphantom{-}
   +
\sqrt{4-\mu ^2} \left(54 \mu ^8-453 \mu ^6+1096 \mu ^4-660 \mu ^2+32
\vphantom{ \left(\mu ^2-4\right)^2\left(\mu ^2 \ln\mu^2\right)}
\right.\nonumber\\
&\hphantom{-+\sqrt{4-\mu ^2}(}
\left.
-3 \left(\mu ^2-4\right)^2 \left(9 \mu ^4-17 \mu ^2+6\right) \mu ^2 \ln\mu^2\right)   \biggr]\nonumber\\
&=3.7~\mathrm{GeV}^{-5}
\,.\label{eq:pllpin}
\end{align}
\paragraph{Neutron}
\begin{align}
P'^{(M1,M1)1}(0)|^{\pi N}_n & =\frac{g_A^2}{288\pi^2 f_\pi^2 M^3}\frac{1}{\mu^2(4-\mu^2)^{3/2}}
\nonumber\\
&\times \biggl[
\mu  \left(3 \mu ^4-18 \mu ^2+10\right) \arccos
\left(\frac{\mu ^2}{2}-1\right)
+\sqrt{4-\mu ^2} \left(8 \mu ^2-4-3\left(\mu ^2-4\right) \mu ^2 \ln
   \mu^2\right)\biggr]\nonumber\\
   &=-2.5~\mathrm{GeV}^{-5} \,,
\displaybreak[0]\label{eq:pmmpin_n}\\
P'^{(L1,L1)1}(0)|^{\pi N}_n & = \frac{g_A^2}{144\pi^2 f_\pi^2 M^3}\frac{1}{\mu^2(4-\mu^2)^{5/2}}
\nonumber\\
&\times
\biggl[
3\mu  \left( \mu ^6-10
   \mu ^4+30 \mu ^2-12\right) \arccos\left(\frac{\mu ^2}{2}-1\right)
   \nonumber\\
&\hphantom{-}
   +
\sqrt{4-\mu ^2} \left(7\mu ^4-50 \mu ^2+16
-3 \left(\mu ^2-4\right)^2  \mu ^2 \ln\mu^2\right)   \biggr]\nonumber\\
&=5.2~\mathrm{GeV}^{-5}
\,.\label{eq:pllpin_n}
\end{align}
\item $\pi\Delta$ loops:
\begin{align}
P'^{(M1,M1)1}(0)|^{\pi \Delta} & =\frac{h_A^2}{31104\pi^2 f_\pi^2 M M^2_\Delta}
\nonumber\\
&\times
\Biggl[-42-24\delta +\int\limits_0^1\mathrm{d} x
\Biggl(\frac{24 (1 - x) x^5 (1 + x + \delta)}{D_\Delta(x)^2}
\nonumber\\
&\hspace{4cm} +\frac{12 x^2 (24 x^3 - 
   x^2 (22 -8 \delta)- 3 x (2 + 5 \delta) + 9 (1 + \delta) )}{D_\Delta(x)}
   \nonumber\\
&\hspace{4cm}-24 x (16 x^2- x (3 - 9 \delta) - 12 (1 + \delta) )
\left[\Xi-\ln D_\Delta(x)\right]\Biggr)\Biggr]\nonumber\\
&=0.5~\mathrm{GeV}^{-5}\,,
\label{eq:pmmpid}\displaybreak[0]
\\
P'^{(L1,L1)1}(0)|^{\pi \Delta} & =\frac{h_A^2}{31104\pi^2 f_\pi^2 M M^2_\Delta}
\nonumber\\
&\times
\Biggl[-16 +\int\limits_0^1\mathrm{d} x
\Biggl(\frac{-24 x^4 (7 - 12 x + 5 x^2) (1 + x + \delta)}{D_\Delta(x)^2}
\nonumber\\
&\hspace{3cm} +\frac{12 x^2 (48 x^3- x^2 (51 - 46 \delta) - 
   x (25 + 71 \delta) + 24 (1 + \delta) )}{D_\Delta(x)}
   \nonumber\\
&\hspace{4cm}-
48 x^2 (7 x-6)\left[\Xi-\ln D_\Delta(x)\right]\Biggr)\Biggr]\nonumber\\
&=-0.8~\mathrm{GeV}^{-5}
\,,\label{eq:pllpid}
\end{align}
with
\begin{equation}
D_\Delta(x)=x^2 + (1 + \delta)^2 - x (2 + 2 \delta + \delta^2 - \mu^2).
\end{equation}
The divergent parts of the polarizabilities, absorbed by higher-order contact terms,
are renormalized according to the modified minimal subtraction ($\overline{\mathrm{MS}}$) scheme, by setting 
to 0 the factor arising in the dimensional regularization: 
\begin{equation}
\Xi=\frac{2}{4-D}-\gamma_E+\ln\frac{4\pi\Lambda^2}{M^2},
\end{equation}
with $D\simeq 4$ the number of dimensions, $\gamma_E$ the Euler constant, and $\Lambda$ the renormalization scale.
\item $\Delta$-pole
\begin{align}
P'^{(M1,M1)1}(0)|^{\Delta\text{-pole}} & =\frac{1}{6M^2 M_+^2}\left(
-\frac{g_M^2}{\varDelta}+2\frac{g_M g_E}{M_+}+\frac{g_E^2}{M_+}
\right)=-1.3~\mathrm{GeV}^{-5}\,,
\\
P'^{(L1,L1)1}(0)|^{\Delta\text{-pole}} & =\frac{1}{6M^2 M_+^2}\left(
-\frac{g_M g_E}{\varDelta}+\frac{g_E^2}{M_+}
\right)=0.4~\mathrm{GeV}^{-5},
\end{align}
with $M_+=M+M_\Delta$, and where $g_M$ ($g_E$) are the M1 (E2) $\gamma N \Delta$ couplings respectively~\cite{Pascalutsa:2006up}.   
\end{enumerate}

We can expand the above expressions in the small scales in order to compare with the
HB expressions given above. Note that these B$\chi$PT results, namely,
the answers for the $\pi N$ loop contribution, do not include the photon coupling to the nucleon anomalous magnetic moment in the loop.
Expanding Eqs.~(\ref{eq:pmmpin})--(\ref{eq:pllpin_n}), we get
\begin{align}
P'^{(M1,M1)1}(0)|^{\pi N}_p & =\frac{g_A^2}{(4\pi f_\pi)^2}\frac{1}{18 M^3\mu^2}
\left[-1+\frac{7\pi}{4}\mu +\left(\frac{45}{2}+18\ln\mu\right)\mu^2+\dots\right]\,,
\\
P'^{(L1,L1)1}(0)|^{\pi N}_p & =\frac{g_A^2}{(4\pi f_\pi)^2}\frac{1}{9 M^3\mu^2}
\left[1-\frac{17\pi}{8}\mu -18\left(1+\ln\mu\right)\mu^2+\dots\right]\,,\\
P'^{(M1,M1)1}(0)|^{\pi N}_n & =\frac{g_A^2}{(4\pi f_\pi)^2}\frac{1}{18 M^3\mu^2}
\left[-1+\frac{5\pi}{4}\mu +\left(\frac{1}{2}+6\ln\mu\right)\mu^2+\dots\right]\,,
\\
P'^{(L1,L1)1}(0)|^{\pi N}_n & =\frac{g_A^2}{(4\pi f_\pi)^2}\frac{1}{9 M^3\mu^2}
\left[1-\frac{9\pi}{8}\mu -\left(\frac{3}{2}+6\ln\mu\right)\mu^2+\dots\right]\,,
\end{align}
which coincides up to the NLO with the result of Eqs.~(\ref{m1m1chpt})--(\ref{l1l1chpt})
if one sets $\kappa_S=\kappa_V=0$. 

The $\pi \Delta$ loop contributions can be expanded in the small quantities
$\delta$ and $\mu$ by, e.g., substituting $\delta\to \eta\delta,\ \mu\to\eta\mu$
into Eqs.~\eqref{eq:pmmpid} and ~\eqref{eq:pllpid}, integrating over
the Feynman parameter $x$, expanding the result in 
powers of $\eta$ and setting $\eta = 1$ in the end. This results in
\begin{align}
P'^{(M1,M1)1}(0)|^{\pi \Delta} & =\frac{h_A^2}{1296\pi^2 f_\pi^2 M M_\Delta^2}
\frac{1}{(\delta^2-\mu^2)}\nonumber\\
&\times
\left[-1+\frac{\delta}{\sqrt{\delta^2-\mu^2}}\ln\frac{\delta+\sqrt{\delta^2-\mu^2}}{\mu}
-\frac{9}{2}\delta+\frac{4\delta^2+5\mu^2}{2\sqrt{\delta^2-\mu^2}}
\ln\frac{\delta+\sqrt{\delta^2-\mu^2}}{\mu}+\dots
\right]
\,,
\\
P'^{(L1,L1)1}(0)|^{\pi \Delta} & =\frac{h_A^2}{648\pi^2 f_\pi^2 M M_\Delta^2}
\frac{1}{(\delta^2-\mu^2)}\nonumber\\
&\times
\left[1-\frac{\delta}{\sqrt{\delta^2-\mu^2}}\ln\frac{\delta+\sqrt{\delta^2-\mu^2}}{\mu}
+\delta-\frac{\delta^2+\mu^2}{2\sqrt{\delta^2-\mu^2}}
\ln\frac{\delta+\sqrt{\delta^2-\mu^2}}{\mu}+\dots
\right]
\,,
\end{align}
whose LO coincides with the LO HB$\chi$PT results of Eqs.~(\ref{sse1}-\ref{sse2}).

\section{Empirical verification}
\label{sec7}

In the following, we will investigate the empirical verification of the sum rules for 
$I_1^\prime(0)$ as well as for $\delta_{LT}$.  
The {\it rhs} of the sum rule of Eq.~(\ref{qsqrgdhsr}) requires the 
information for the Pauli radius and phenomenological dispersive estimates for the 
spin and generalized spin polarizabilities.  
To evaluate the Pauli radius, we can use the experimental information on the electric $\langle r_E^2 \rangle$ 
and magnetic $\langle r_M^2 \rangle $ radii of the nucleon. 
The Pauli radius is obtained from those quantities as
\begin{eqnarray} 
\kappa_N \langle r_2^2 \rangle = \mu_N \langle r_M^2 \rangle -  \langle r_E^2 \rangle + \frac{3 \kappa_N}{2 M^2}, 
\end{eqnarray}
where $\mu_N$ denotes the nucleon magnetic moment. Using the recent experimental values for the proton electric and magnetic radii from Ref.~\cite{Bernauer:2013tpr},
\begin{eqnarray}
\langle r_E^2 \rangle &=& \, 0.77 \pm 0.01 \; {\mathrm {fm}}^2 , \nonumber \\
\langle r_M^2 \rangle &=& \, 0.60 \pm 0.03 \; {\mathrm {fm}}^2 , 
\end{eqnarray}
one obtains for the proton Pauli radius
\begin{eqnarray}
\langle r_2^2 \rangle = 0.58 \pm 0.04 \;  {\mathrm {fm}}^2. 
\end{eqnarray}

We next turn to the spin polarizability contributions to both sum rules of Eqs.~(\ref{qsqrgdhsr}) 
and (\ref{S2sr3}). These spin polarizabilities contain in general a $\pi^0$-pole contribution and a non-pole contribution. The latter can be evaluated by an unsubtracted dispersion relation.
The $\pi^0$-pole contribution to the relevant spin polarizabilities and to the slopes of the spin GPs entering the sum rule are given by
\begin{eqnarray}
&&\frac{1}{\alpha_\mathrm{em}}  \gamma_{E1 M2} |^{\pi^0}= \frac{1}{2}  C_\pi , \quad \quad
\frac{1}{\alpha_\mathrm{em}}  \gamma_{E1 E1} |^{\pi^0}= \frac{1}{2}  C_\pi , \nonumber \\
&&P^{\prime \, (M1, M1)1}(0) |^{\pi^0} = \frac{1}{6 M} C_\pi , \quad \quad
P^{\prime \, (L1, L1)1}(0) |^{\pi^0} = 0,  
\end{eqnarray}
with $C_\pi = g_A / (4 \pi^2 f_\pi^2 m_\pi^2)$. By inserting these into the sum rules of Eqs.~(\ref{qsqrgdhsr}) and (\ref{S2sr3}), one 
notices that the $\pi^0$-pole contribution drops out of the {\it rhs} of both sum rules.  This is consistent, as the {\it lhs} of these sum rules, corresponding with the moments of $g_1$ and $g_2$, do not contain such $\pi^0$-pole contributions.  

The non-pole parts of the spin polarizabilities have been estimated phenomenologically using unsubtracted dispersion relations~\cite{Holstein:1999uu, Drechsel:2002ar}. The corresponding dispersive estimates for the 
generalized polarizabilities have been performed in Refs.~\cite{Pasquini:2001yy, Drechsel:2002ar}.  
To show the uncertainty due to the phenomenological input in the dispersion relations, we show in Table~\ref{sumruletest} the dispersive results for the spin and generalized spin polarizabilities using 
either MAID2000~\cite{Drechsel:1998hk}  or 
 MAID2007~\cite{Drechsel:2007if} as input for the $\pi N$ channel contribution. 

Very recently, a first experimental extraction of the four proton spin polarizabilities was performed using 
polarized Compton scattering on a proton target, resulting in the values~\cite{Martel:2014pba}
\begin{eqnarray}
\gamma_{E1 M2} &=& (- 0.7 \pm 1.2 ) \times 10^{-4} \, {\mathrm{fm}}^4  , \\ 
\gamma_{E1 E1} &=& (- 3.5 \pm 1.2 ) \times 10^{-4} \, {\mathrm{fm}}^4  . 
\end{eqnarray}
We like to notice that the above values result from a fit of the four proton dipole spin polarizabilities to one double polarization Compton scattering observable, one single polarization observable (photon asymmetry), the backward spin polarizability combination $\gamma_\pi$, extracted from unpolarized experiments, as well as the forward spin polarizability combination $\gamma_0$. 
The large error of $\gamma_{E1 M2}$ which results from the fit in Ref.~\cite{Martel:2014pba} is mainly due to the present large error on $\gamma_\pi$, given by $\delta \gamma_\pi = 1.8  \times 10^{-4} \, {\mathrm{fm}}^4 $. Ongoing measurements of another double polarization Compton scattering observable will allow one to reduce the error on $\gamma_\pi$ by a factor of 4, 
which is expected to reduce the error on the other spin polarizabilities accordingly.

\begin{table}[h]
{\centering 
\begin{tabular}{|c|c|c|c|c|c|c|}
\hline
&  Disp. rel. & Disp. rel. & HB$\chi$PT & B$\chi$PT & B$\chi$PT & Experiment \\
&  MAID2000 & MAID2007& $\pi N$ to ${\mathcal O}(p^4)$ & $\pi N$  & $\pi N + \Delta + \pi \Delta$ &\\
\hline
\hline
$\gamma_{E1 M2}$ 
& $-0.03$  \quad \cite{Holstein:1999uu}  & $-0.1$  \cite{Drechsel:2002ar,Drechsel:2007if}  
& $0.2$ \cite{VijayaKumar:2000pv} 
& $0.5$ \cite{Lensky:2015awa}
& $0.2 \pm 0.2$ \cite{Lensky:2015awa}
& $-0.7 \pm 1.2$   \cite{Martel:2014pba}   \\
\, [$10^{-4}$ fm$^4$]  & & &  & & & \\
\hline
$\gamma_{E1 E1}$   
& $-4.3$  \quad \cite{Holstein:1999uu} 
& $-4.3$   \cite{Drechsel:2002ar,Drechsel:2007if}  
& $-1.3$ \cite{VijayaKumar:2000pv} 
&  $-3.4$ \cite{Lensky:2015awa}
&  $-3.3 \pm 0.8$ \cite{Lensky:2015awa}
& $-3.5 \pm 1.2$   \cite{Martel:2014pba}   \\
\, [$10^{-4}$ fm$^4$] & & &  & & & \\
\hline
$P^{\prime \, (M1, M1)1}(0)$  
& $-5.8$  \cite{Pasquini:2001yy,Drechsel:1998hk}   
& $-4.7$  \cite{Pasquini:2001yy,Drechsel:2007if}  
& $-0.7$ \cite{Kao:2002cn,Kao:2004us}
& $-2.3$
& $-3.0 \pm 0.7$ \cite{Lensky:2016nui}
& $-$ \\
\, [GeV$^{-5}$]  & & &  & & & \\
\hline
$P^{\prime \, (L1, L1)1}(0)$   
& \quad $3.4$  \cite{Pasquini:2001yy,Drechsel:1998hk}  
& \quad $4.3$  \cite{Pasquini:2001yy,Drechsel:2007if} 
& $-1.3$ \cite{Kao:2002cn, Kao:2004us}
& $3.7$
& $3.4 \pm 0.9$ \cite{Lensky:2016nui}
& $-$ \\
\, [GeV$^{-5}$]  & & & & & &\\
\hline
\hline
$ \langle r_2^2 \rangle$ & &  & $0.55$ \cite{BKM95} & $-$ & $-$
& $ 0.58 \pm 0.04 $  \cite{Bernauer:2013tpr}    \\
 \, [fm$^2$]  &  & &  & &  &\\
\hline
\quad $I_1^\prime(0)$  
&  \quad $6.8$  (SR)
& \quad $4.0$  (SR) 
& 7.1 \cite{Ji:1999pd, Ji:1999mr}
& $-$ 
& $-$ 
&   $7.6 \pm 2.5$  \cite{Prok:2008ev}    \\
\, [GeV$^{-2}$]  & & & & & & \\
\hline
\quad $I_1^\prime(0) - \frac{\kappa_N^2}{12} \langle r_2^2 \rangle$  
&  \quad $2.9$  (SR)
& \quad $0.1$  (SR) 
& 3.3 \cite{Ji:1999pd, Ji:1999mr}
& $0.3$
& $0.3 \pm 0.1$ \cite{Lensky:2014dda}
&   $2.6 \pm 2.5$  \cite{Prok:2008ev}    \\
\, [GeV$^{-2}$]  & & & & & & \\
\hline
\quad $\delta_{LT}$   
&  \quad $1.5$    (SR )
&  \quad $1.5$  (SR)
& $1.5$ \cite{Kao:2002cp}
& $1.5$ 
& $1.4 \pm 0.3$ \cite{Lensky:2014dda}
&   $1.34$ \cite{Drechsel:2000ct,Drechsel:2007if}   \\
\, [$10^{-4}$ fm$^4$]   & & & & & & (MAID2007)    \\
\hline
\end{tabular}\par
}
\caption{Estimates for the different quantities which enter the generalized GDH sum rule of Eq.~(\ref{qsqrgdhsr}) and the sum rule for $\delta_{LT}$ of Eq.~(\ref{S2sr3}) for the proton. 
The first  and second data columns show the dispersive estimates (Disp. rel.) using respectively MAID2000~\cite{Drechsel:1998hk}  and MAID2007~\cite{Drechsel:2007if} as input for the (generalized) spin polarizabilities. 
Third data column: HB$\chi$PT results to ${\mathcal O}(p^4)$. 
Fourth data column: $\pi N$ loop results within covariant B$\chi$PT.
Fifth data column: $\pi N$ loop + $\Delta$-pole + $\pi \Delta$ loop results within covariant B$\chi$PT (with errors calculated as explained in Ref.~\cite{Lensky:2015awa}).
The last column shows the experimental values. 
The values for $I_1^\prime(0)$ and $\delta_{LT}$ show the corresponding SR estimates, based on Eqs.~(\ref{qsqrgdhsr}) and (\ref{S2sr3}) respectively. 
We also show the result for the quantity $I_1^\prime(0) - \frac{\kappa_N^2}{12} \langle r_2^2 \rangle$, which is predictable in B$\chi$PT.  
}
\label{sumruletest}
\end{table}

Based on the above experimental and phenomenological values, 
we compare in Table~\ref{sumruletest} the different contributions to the proton generalized GDH sum rule of Eq.~(\ref{qsqrgdhsr}) as well as the sum rule of Eq.~(\ref{S2sr3}) for $\delta_{LT}$. 
For the sum rule of Eq.~(\ref{qsqrgdhsr}), we can compare this result directly with the 
experimental value for $I_1^\prime(0)$ as measured by JLab/CLAS~\cite{Prok:2008ev}. 
We see that within the error bars, the phenomenological DR estimate for the 
proton sum rule (SR) value of $I_1^\prime(0)$ is in good agreement with the experimental value. 
For the sum rule of Eq.~(\ref{S2sr3}), the experimental value of $\delta_{LT}$ is not yet 
available. Comparing with the phenomenological estimate of Ref.~\cite{Drechsel:2000ct}, 
one finds an agreement of this sum rule within 10\%; see the last row in Table~\ref{sumruletest}. 
A recent JLab experiment to measure $\delta_{LT}$, which is currently under analysis, will allow one to provide a direct experimental verification of the sum rule of Eq.~(\ref{S2sr3}) in the near future. 

\begin{figure}
\begin{center}
\includegraphics[width=0.6\textwidth]{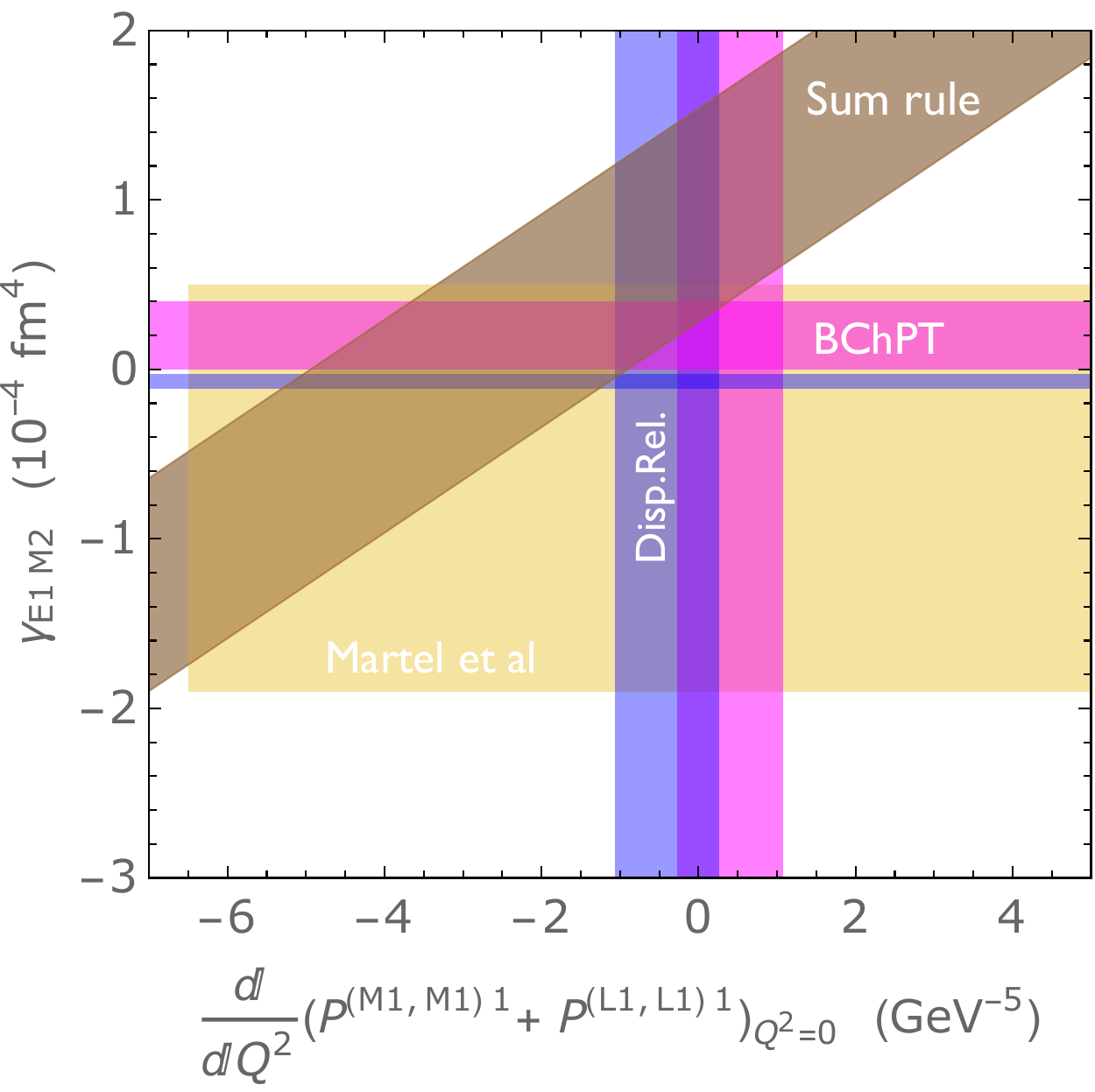}
\caption{The sum rule relation of Eq.~(\ref{qsqrgdhsr}) between  
$\gamma_{E1 M2}$ versus $\left( P^{\prime \, (M1, M1)1}(0) + P^{\prime \, (L1, L1)1}(0) \right)$. The brown band is the sum rule constraint based on the empirical information for $I_1^\prime(0)$ from~\cite{Prok:2008ev}, and for $ \langle r_2^2 \rangle$ 
from~\cite{Bernauer:2013tpr}. The yellow band is the empirical extraction of 
$\gamma_{E1 M2}$ from~\cite{Martel:2014pba}. 
The purple bands are the DR evaluations~\cite{Drechsel:2002ar} for the RCS and VCS polarizabilities, where the width of the bands is obtained by using either 
MAID2000~\cite{Drechsel:1998hk} or MAID2007~\cite{Drechsel:2007if} as input in the dispersive evaluations. 
The pink bands are the B$\chi$PT evaluations  
($\pi N + \Delta + \pi \Delta$ results from Table~\ref{sumruletest})~\cite{Lensky:2014dda,Lensky:2015awa,Lensky:2016nui}. }
\label{Fig:across}
\end{center}
\end{figure}

\begin{figure}
\begin{center}
\includegraphics[width=0.6\textwidth]{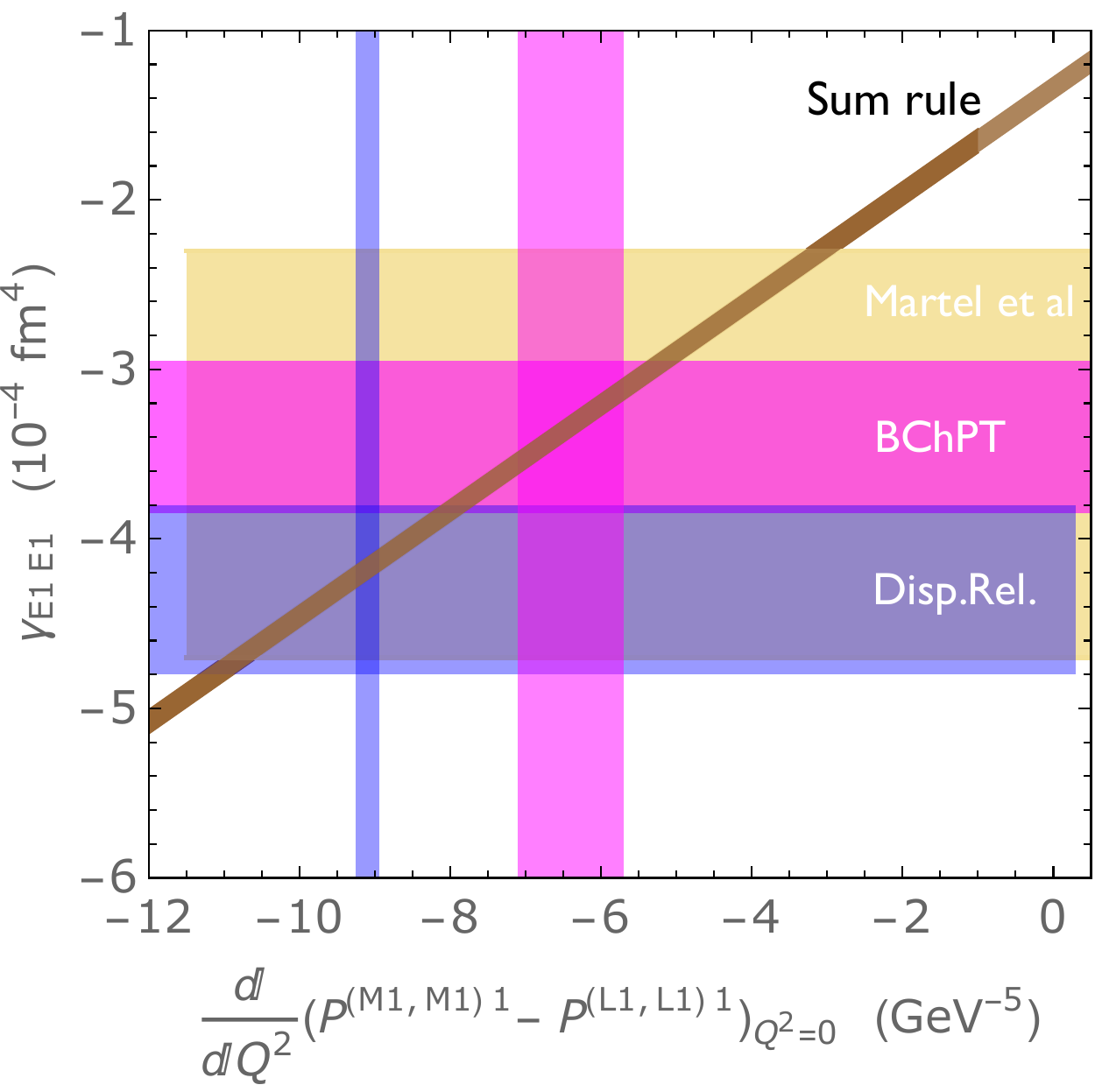}
\caption{The sum rule relation of Eq.~(\ref{S2sr3}) between 
$\gamma_{E1 E1}$ versus $\left( P^{\prime \, (M1, M1)1}(0) - P^{\prime \, (L1, L1)1}(0) \right)$. 
The brown band is the sum rule constraint based on the phenomenological MAID2007~\cite{Drechsel:2007if}  information for $\delta_{LT}$. 
The yellow band is the empirical extraction of 
$\gamma_{E1 E1}$ from~\cite{Martel:2014pba}. 
The purple bands are the DR evaluations~\cite{Drechsel:2002ar} for the RCS and VCS polarizabilities, where the width of the bands is obtained by using either 
MAID2000~\cite{Drechsel:1998hk} or MAID2007~\cite{Drechsel:2007if} as input in the dispersive evaluations. 
The pink bands are the B$\chi$PT evaluations  
($\pi N + \Delta + \pi \Delta$ results from Table~\ref{sumruletest})~\cite{Lensky:2014dda,Lensky:2015awa,Lensky:2016nui}.
}
\label{Fig:across2}
\end{center}
\end{figure}

We provide a graphical presentation of both spin-dependent sum rules in 
Figs.~\ref{Fig:across} and \ref{Fig:across2}. 
Using only the empirical information for $I_1^\prime(0)$ and $\delta_{LT}$, the sum rules provide a slanted (brown) band in the plots of $\gamma_{E1M2}$ and  $\gamma_{E1E1}$  
versus the slopes of the GPs.
The pioneering experimental values for $\gamma$'s, recently obtained by the A2 Collaboration at MAMI~\cite{Martel:2014pba}, are shown by the broad horizontal (yellow) band in the figures. The region where the two bands overlap yields a prediction for the slopes of the GPs. A measurement of GP slopes using VCS is required to directly verify this prediction. 
One sees from both figures that the phenomenological DR estimates 
of Pasquini {\em et al.}~\cite{Drechsel:2002ar} (shown by the horizontal and vertical purple bands)
are well in agreement, within uncertainties, with the RCS spin polarizabilities and are consistent with the sum rule bands. 
The figures also show the results obtained in the covariant B$\chi$PT 
($\pi N + \Delta + \pi \Delta$).  We have checked above that both the covariant
and HB$\chi$PT calculations satisfy the sum rules exactly.
Within the present error bars, the B$\chi$PT results are in agreeement with the RCS
spin polarizabilities [the same is also true for the HB$\chi$PT calculation (not shown
in the figures), although for $\gamma_{E1 E1}$ the HB$\chi$PT extraction yields a large uncertainty~\cite{Griesshammer:2015ahu}].
The B$\chi$PT results for the slopes of the two spin GPs are also in  good agreement
with the DR estimates, as noted in Ref.~\cite{Lensky:2016nui}.

\begin{figure}
    \centering        \includegraphics[width=0.6\textwidth]{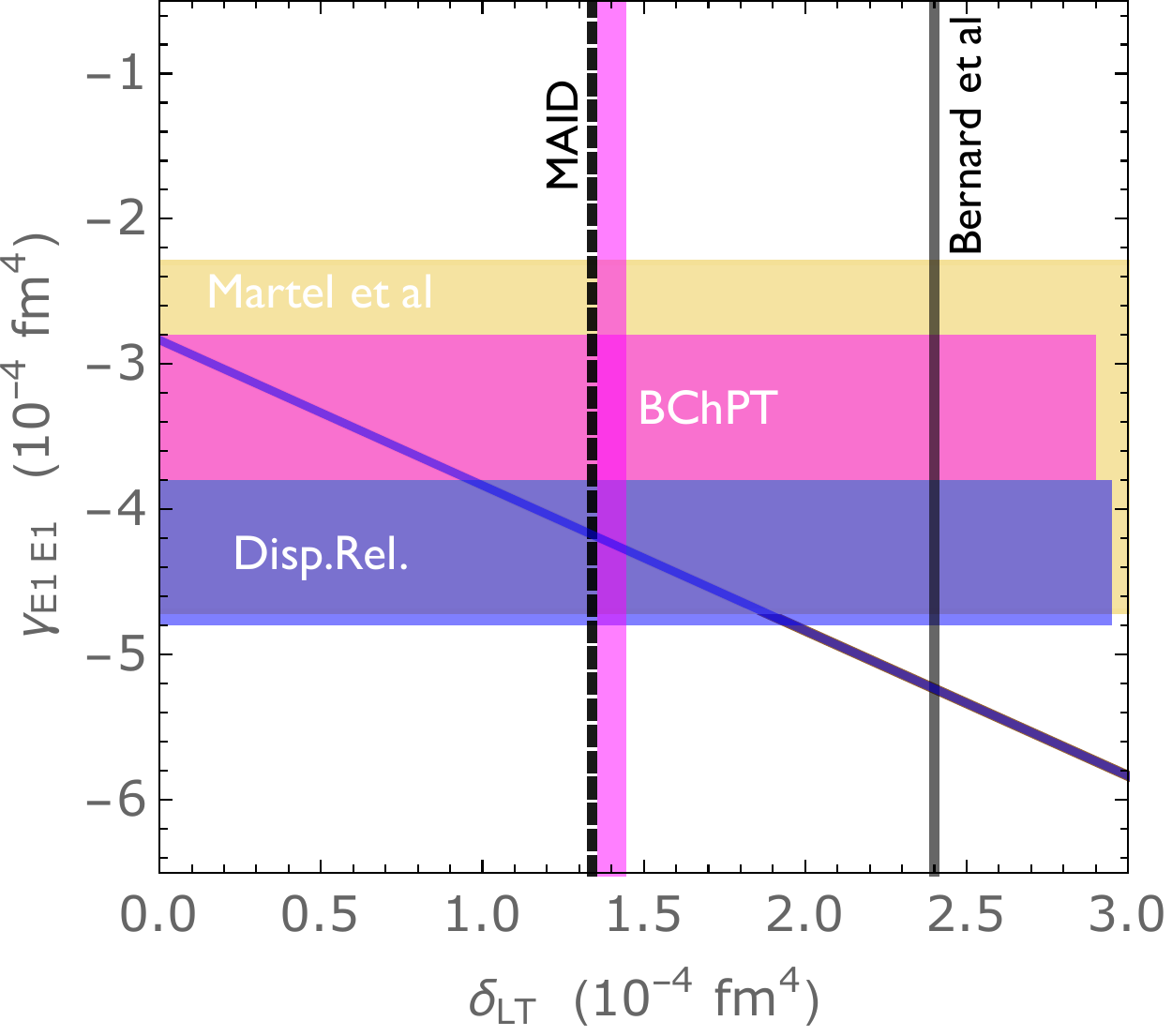}
                     \caption{Spin polarizabilities,
                     $\gamma_{E1E1}$ versus $\delta_{LT}$,  for the proton.  Results for 
 $\gamma_{E1E1}$ (horizontal bands) are from: the experiment of \cite{Martel:2014pba}  (yellow), the B$\chi$PT calculation of \cite{Lensky:2015awa} (red),
 and the fixed-$t$ DR calculation 
 of \cite{Holstein:1999uu, Drechsel:2002ar} (purple). Results for 
  $\delta_{LT}$ (vertical bands) are from:
  MAID2007 \cite{Drechsel:2000ct,Drechsel:2007if}  (dashed line), 
  \cite{Lensky:2014dda} (pink), and \cite{Bernard:2012hb} (gray).
  The line across is based on the sum rule
   using the values of GPs from
  the DR calculation of \cite{Pasquini:2001yy}.
  \label{fig:acrossSR}}
\end{figure}

In Fig.~\ref{fig:acrossSR}, we provide an alternative presentation of the sum rule of 
Eq.~(\ref{S2sr3}) by presenting $\gamma_{E1 E1}$ versus $\delta_{LT}$. The value of the spin GP combination $\left( P^{\prime \, (M1, M1)1}(0) - P^{\prime \, (L1, L1)1}(0) \right)$ is taken from the DR estimate, yielding the slanted (purple) sum rule band. For the spin polarizabilities, we have presented two variants of covariant B$\chi$PT: the results of 
Lensky et al.~\cite{Lensky:2014dda, Lensky:2015awa}  shown in Table~\ref{sumruletest}, 
and the results of Bernard et al.~\cite{Bernard:2012hb}. They are done in two different counting schemes for the $\Delta$-isobar contribution. One notices that they yield a noticeable difference in the value for $\delta_{LT}$ which is mainly due to a much larger contribution of the $\pi \Delta$ loops in Ref.~\cite{Bernard:2012hb} as compared to Ref.~\cite{Lensky:2014dda}. One notices that the phenomenological MAID estimate~\cite{Drechsel:2000ct,Drechsel:2007if} for $\delta_{LT}$ favors the smaller value for $\delta_{LT}$ of both B$\chi$PT variants. The recent JLab proton $\delta_{LT}$ experiment, which is currently under analysis, will allow a direct experimental verification of this puzzle.    

\section{Predictions for the VCS response function $P_{TT}$ at low $Q^2$}
\label{sec8}

One combination of the VCS spin polarizabilities can be obtained in an unpolarized 
VCS experiment. 
At low energy ($\rmqp$) of the emitted photon, the energy dependence  
of the $e p \to e p \gamma$ unpolarized cross section can be expressed 
through a Taylor expansion in $\rmqp$, taking the lowest three terms into account. 
In such an expansion in $\rmqp$, the experimentally extracted 
VCS unpolarized squared amplitude 
 $\calm^{\rm exp}$ takes the form \cite{Guichon:1995pu}
\begin{equation}
\calm^{\rm exp}=\frac{\calm^{\rm exp}_{-2}}{\rmqp^2}
+\frac{\calm^{\rm exp}_{-1}}{\rmqp}
+\calm^{\rm exp}_0+O(\rmqp) \, .
\label{eq:unpolsqramp}
\end{equation}
Due to the low-energy theorem (LET), the threshold coefficients 
$\calm^{\rm exp}_{-2}$ and $\calm^{\rm exp}_{-1}$ are
known~\cite{Guichon:1995pu}, and are fully determined from the Bethe-Heitler +
Born (BH + Born) amplitudes. 
The information on the GPs is contained in \( \calm^{\rm exp}_0\), 
which contains a part originating from the 
BH+Born amplitudes and another one which is a linear combination of the GPs, with
coefficients determined by the kinematics. 
The unpolarized observable $\calm^{\rm exp}_0$ 
can be expressed in terms of three structure functions
$P_{LL}(Q^2)$, $P_{TT}(Q^2)$, and $P_{LT}(Q^2)$ by ~\cite{Guichon:1995pu}
\begin{eqnarray}
\hspace{-0.7cm}
\calm^{\rm exp}_0 - \calm^{\rm BH+Born}_0 
= 2 K  \Bigg\{ 
v_1 \left[ {\varepsilon  P_{LL}(Q^2) - P_{TT}}(Q^2)\right] 
+ \left(v_2 + \sqrt{\frac{\tau}{1 + \tau}} v_3\right)\sqrt {2\varepsilon \left( 
{1+\varepsilon }\right)} P_{LT}(Q^2) \Bigg\}, 
\label{eq:vcsunpol}
\end{eqnarray}
where $K$ is a kinematical factor, $\varepsilon$ is the virtual
photon polarization (in the standard notation used in electron
scattering), and $v_1, v_2, v_3$ are kinematical
quantities depending on $\varepsilon$ and $Q^2$  
as well as on the c.m.\ polar and azimuthal angles of the
produced real photon (for details see Ref.~\cite{Guichon:1998xv}). 
The three unpolarized observables of Eq.~(\ref{eq:vcsunpol}) 
can be expressed in terms of the six GPs as \cite{Guichon:1995pu,Guichon:1998xv}
\begin{eqnarray}
&&P_{LL} \,=\, - 2\sqrt{6} \, M \, G_E \, P^{\left( {L1,L1} \right)0} \;, 
\label{eq:unpolobsgp1} \\
&&P_{TT}=\hphantom{-}6 M G_M(1+\tau)
\left[
2\sqrt{2}\,M \tau\, P^{(L1,M2)1}+P^{(M1,M1)1}
\right]
\,,\label{eq:unpolobsgp2} \\
&&P_{LT}=\hphantom{-}\sqrt{\mbox{$\frac{3}{2}$}}M\sqrt{1+\tau}
\left[
G_E P^{(M1,M1)0}-\sqrt{6}\,G_M P^{(L1,L1)1}
\right],
\label{eq:unpolobsgp3}
\end{eqnarray}
where the nucleon form factors $G_E\equiv F_D+\tau F_P$ and $G_M$ depend on $Q^2$. 

We notice from Eq.~(\ref{eq:unpolobsgp2}) that the VCS response function $P_{TT}$ 
involves only spin GPs and vanishes at $Q^2 = 0$. 
The slope at $Q^2 = 0$ of $P_{TT}$ can be expressed as
\bea
P^\prime_{TT}(0) &\equiv& \frac{{\rm d}\hphantom{Q^2}}{{\rm d} Q^2}  P_{TT}(Q^2) \bigg|_{Q^2 = 0} \nonumber \\
&=& 2 \mu_N \left\{ 3 M  P^{\prime \, (M1, M1)1}(0) - \frac{1}{\alpha_\mathrm{em}}  \gamma_{E1 M2} \right\} \nonumber \\
&=& 2 \mu_N \left\{ 
\frac{1}{M^2} \left( \frac{ \kappa_N^2}{12} \langle r_2^2 \rangle  - I_1^\prime(0) \right) +  
\frac{1}{2 \alpha_\mathrm{em}} \left(- \gamma_{E1 M2} + \gamma_{E1 E1} + \delta_{LT} \right) 
\right\},
\label{eq:Ptt}
\eea
with $\mu_N$ the nucleon magnetic moment, and where the last line has been obtained by 
eliminating $P^{\prime \, (M1, M1)1} (0)$ by using the sum rules of Eqs.~(\ref{qsqrgdhsr}) and (\ref{S2sr3}).  

When plugging the respective values into the last line of Eq.~(\ref{eq:Ptt}), 
by using the values listed in Table~\ref{sumruletest}, we obtain the DR prediction, the respective $\chi$PT predictions, as well as the empirical prediction for the slope at $Q^2 = 0$ of $P_{TT}$, 
which we list in Table~\ref{ptttab}. 

\begin{table}[h]
{\centering 
\begin{tabular}{|c|c|c|c|c|c|c|}
\hline
&  Disp. rel. & Disp. rel. & HB$\chi$PT & \quad B$\chi$PT \quad &  B$\chi$PT & \quad Empirical \quad \\
&  MAID2000 & MAID2007& $\pi N$ to ${\mathcal O}(p^4)$ & $\pi N$  & $\pi N + \Delta + \pi \Delta$ & \quad SR evaluation \quad \\
\hline
\hline
$P^{\prime}_{TT}(0)$  
& $-88$  
& $-68$  
& $-21$ 
& $-63$
& $-60 \pm 10$
& $-53 \pm 46$ \\
\, [GeV$^{-4}$]  & & &  & & & \\
\hline
\end{tabular}\par
}
\caption{Different estimates for the VCS response function $P^{\prime}_{TT}(0)$.}
\label{ptttab}
\end{table}

 \begin{figure}
\begin{center}
\includegraphics[width=0.7\textwidth]{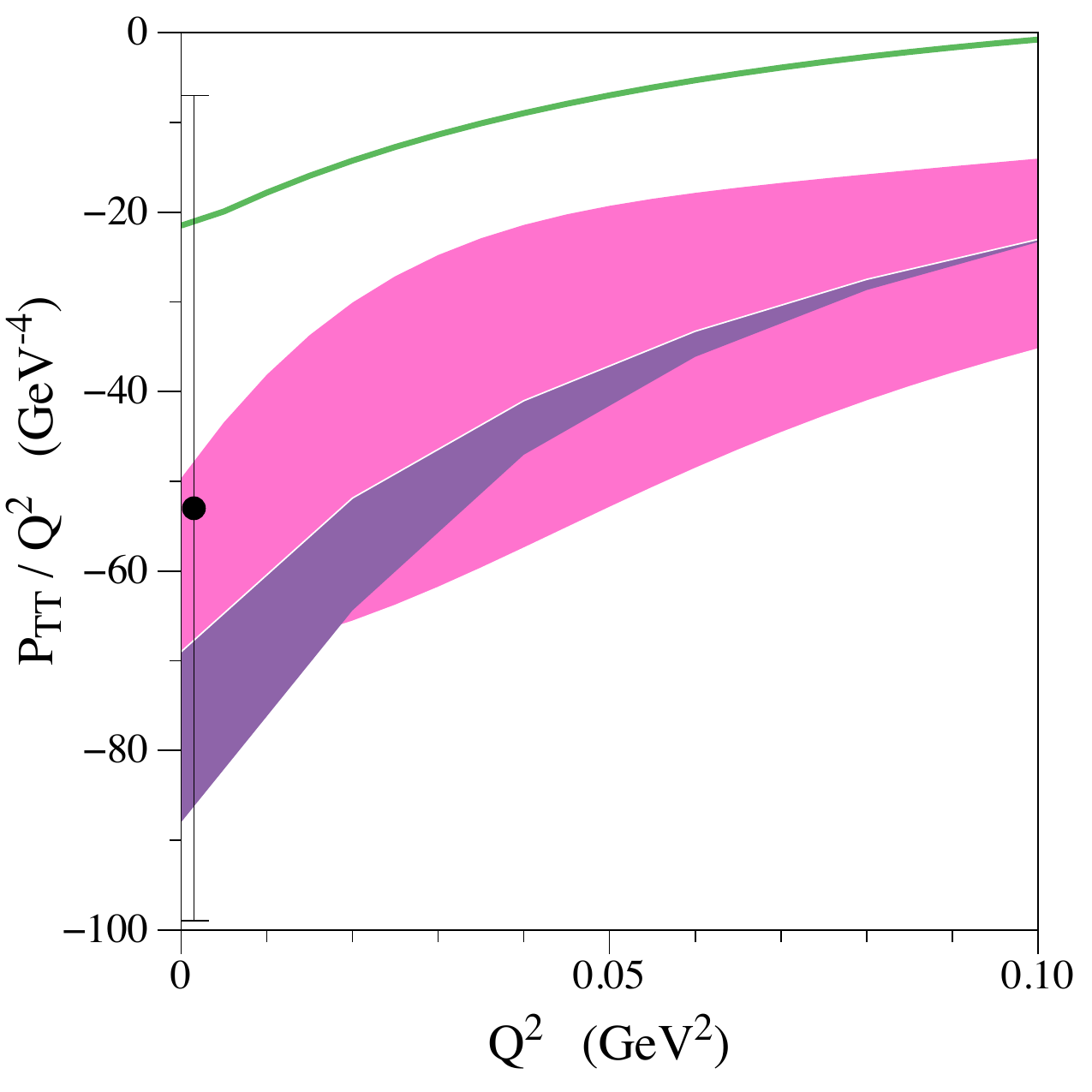}
\caption{$Q^2$ dependence of the VCS response function $P_{TT} / Q^2$. The purple band is the DR estimate~\cite{Drechsel:2002ar} where the width of the band corresponds to the evaluations obtained by using MAID2007~\cite{Drechsel:2007if} (upper) or MAID2000~\cite{Drechsel:1998hk} (lower). The green curve is the NLO prediction of HB$\chi$PT, corresponding with $\pi N$ loops to ${\mathcal O}(p^4)$. 
The red band at low $Q^2$ is the B$\chi$PT prediction ($\pi N + \Delta + \pi \Delta$) of Ref.~\cite{Lensky:2016nui}. The data point at $Q^2 = 0$ corresponds with the empirical sum rule evaluation according to the last column in Table~\ref{ptttab}.}
\label{fig:ptt}
\end{center}
\end{figure}

In Fig.~\ref{fig:ptt}, we show 
the predictions of the DR and chiral calculations for $P_{TT}/Q^2$ 
at low $Q^2\le 0.1$~GeV${^2}$, together with the empirical evaluation at $Q^2 = 0$. 
One can again see that the DR and B$\chi$PT
predictions agree quite well, whereas the HB$\chi$PT curve is much smaller. 
All these theoretical calculations are compatible
with the result of the empirical evaluation within the present sizable uncertainty
of the latter.

\section{Summary and Conclusion}
\label{sec9}

By generalizing the Gerasimov-Drell-Hearn sum rule to finite photon
virtuality, we obtain the two new model-independent
relations. They link the parameters characterizing different sectors of low-energy interactions between the nucleon spin structure and electromagnetic
waves. The parameters, involved in these relations, are extracted from
experimental information on nucleon Compton scattering in different regimes:
RCS (spin polarizabilities), VCS (generalized polarizabilities),
and VVCS (longitudinal-transverse polarizability and the generalized GDH integral). In addition, they involve 
the nucleon form factors in the form of the Pauli radii and
the anomalous magnetic moments. 

These relations are identically verified in B$\chi$PT and in HB$\chi$PT.
We have also studied their empirical consequences, and  found that
the current experimental extractions and
phenomenological estimates done in the fixed-$t$ 
DR framework for the proton 
are consistent with
the sum rules. The B$\chi$PT predictions are also
in agreement with these relations (with the notable exception of the $\delta_{LT}$ where
there appears to be a disagreement between the  B$\chi$PT calculations
of Ref.~\cite{Lensky:2014dda} and  Ref.~\cite{Bernard:2012hb}). 
We have used the relations to evaluate the slope of the VCS response function
$P_{TT}$ at zero virtuality  and compared it with the results of
the DR and of the chiral calculations. 
The covariant B$\chi$PT and the DR give similar results for
$P_{TT}^\prime(0)$, whereas the HB$\chi$PT value is considerably
different from them. The empirical result, obtained using the new
relation has yet a large uncertainty, but in the future
will be able to discriminate between the predictions.

The  new relations have thus been  shown to hold
in  a quantum-field-theoretic  framework and are proving
to be useful in constraining the low-energy spin structure
of the nucleon.

\acknowledgments

We would like to thank Barbara Pasquini for helpful discussions.  
This work was supported by the Deutsche Forschungsgemeinschaft (DFG) 
in part through the Collaborative Research Center [The Low-Energy Frontier of the Standard Model (SFB 1044)], and in part through the Cluster of Excellence [Precision Physics, Fundamental
Interactions and Structure of Matter (PRISMA)], and by the Ministry of Science and Technology of Taiwan under Grants NSC 102-2112-M-033-005-MY3 and MOST 105-2112-M-033-004.

\end{document}